\documentstyle[fleqn,run2col,epsf,amssymb,citesort]{article}
\catcode`@=11
\def\citer{\@ifnextchar
[{\@tempswatrue\@citexr}{\@tempswafalse\@citexr[]}}

\def\@citexr[#1]#2{\if@filesw\immediate\write\@auxout{\string\citation{#2}}\fi
  \def\@citea{}\@cite{\@for\@citeb:=#2\do
    {\@citea\def\@citea{--\penalty\@m}\@ifundefined
       {b@\@citeb}{{\bf ?}\@warning
       {Citation `\@citeb' on page \thepage \space undefined}}%
\hbox{\csname b@\@citeb\endcsname}}}{#1}}
\catcode`@=12
\catcode`@=11
\def\@maketitle{%                 % Actual formatting of \maketitle
  \global\setbox\fm@box=\vbox\bgroup
    \vskip 8mm                    % 930715: 8mm white space above 
{ \begin{flushright} \@preprint \end{flushright}  \vskip\@bls }     %%%%%%%%%%%%% MODIFIED: FIO 
    \raggedright                  % Front-matter text is ragged right
    \hyphenpenalty\@M             % and is not hyphenated.
    {\Large \@title \par}         % Title set in larger font.
    \vskip\@bls                   % One line of vertical space after title.
    {\normalsize                  % each author set in the normal
     \@author \par}               % typeface size
    \vskip\@bls                   % One line of vertical space after author(s).
    \@address                     % all addresses
  \egroup
  \twocolumn[%                    % Front-matter text is over 2 columns.
    \unvbox\fm@box                % Unwrap contents of front-matter box
    \vskip\@bls                   % add 1 line of vertical space,
    \unvbox\abstract@box          % unwrap contents of abstract boxes,
    \vskip 2pc]}                  % and add 2pc of vertical space
\def\preprint#1{\def\@preprint{{#1}}}
\preprint{}
\catcode`@=12
\newcommand{\pr}[3]{{Phys.\ Rev.} {\bf #1} (19#2) #3}
\newcommand{\np}[3]{{Nucl.\ Phys.} {\bf #1} (19#2)~#3}
\newcommand{\pl}[3]{{Phys.\ Lett.} {\bf #1} (19#2) #3}
\newcommand{\beq}{\begin{equation}}
\newcommand{\eeq}{\end{equation}}
\newcommand{\beqa}{\begin{eqnarray}}
\newcommand{\eeqa}{\end{eqnarray}}

\def\simgt{\rlap{\lower 3.5 pt \hbox{$\mathchar \sim$}} \raise 1pt \hbox{$>$}}
\def\simlt{\rlap{\lower 3.5 pt \hbox{$\mathchar \sim$}} \raise 1pt \hbox{$<$}}
\def\eq#1{Eq.~(\ref{eq:#1})}

\def\fig#1{Fig.~\ref{fig:#1}}
\def\tab#1{Table.~\ref{tab:#1}}
\def\junk#1{}
\def\nub{\bar{\nu}}
\def\ptmax{P_{T_{max}}}
\def\sec#1{Sec.~[\ref{sec:#1}]}
\def\bibtmp#1{\bibitem{#1}}

\def\tableht{
\begin{table}[htbp]
\begin{center}
\begin{tabular}{|c|c|c|c|} \hline
$x$ & LO & NLO & NNLO \\ \hline 
0 -- 0.0005 &    $-$0.4754 &   0.0116 &   $-$0.0061 \\
0.0005 -- 0.005 &    $-$0.2512 &   $-$0.0475 &   0.0437 \\
0.005 -- 0.01 &     $-$0.2481 &   $-$0.1376 &    $-$0.0048 \\
0.01 -- 0.06 &     $-$0.2306 &    $-$0.1271 &   $-$0.0359 \\
0.06 -- 0.1 &    $-$0.1373 &   $-$0.0321 &     0.0167 \\
0.1 -- 0.2 &    $-$0.1263 &    $-$0.0361 &    0.0075 \\
0.2 -- 0.3 &    $-$0.1210 &   $-$0.0893 &    $-$0.0201 \\
0.3 -- 0.4 &    $-$0.0909 &   $-$0.1710 &    $-$0.1170 \\
0.4 -- 0.5 &      0.1788 &    $-$0.0804 &    $-$0.0782 \\
0.5 -- 0.6 &      0.8329 &    0.3056 &    0.1936 \\
0.6 -- 0.7 &      2.544  &    1.621  &    1.263  \\
0.7 -- 0.8 &      6.914  &    5.468  &    4.557 \\
0.8 -- 0.9 &     19.92   &   18.03   &   15.38  \\ \hline
\end{tabular}
\caption{Values of the higher-twist coefficient $D_2(x)$ extracted from the LO,
NLO and NNLO fits. Units are GeV$^2$. Table taken from Ref.~\protect\cite{thorne}.
}
\end{center}
\end{table}
}
\def\tablek{
\begin{table}[hbtp]
\begin{center}
\begin{tabular}{|c|c|c|c|}
\hline \hline   %%%%%%%%%%%%%%%%%%%%%%%%%%%%%%%%
Experiment & Order & $\kappa$ & Ref. \\
\hline
\hline  %%%%%%%%%%%%%%%%%%%%%%%%%%%%%%%%
CDHS
& LO 
& $ 0.47 \pm {0.08} \pm {0.05} $
& \cite{cdhsw}
\\ \hline  %%%%%%%%%%%%%%%%%%%%%%%%%%%%%%%%
FMMF
& LO 
& $ 0.41   ^{+0.075}_{-0.075}\; ^{+0.103}_{-0.069}\; $
& \cite{fmmf} 
\\ \hline  %%%%%%%%%%%%%%%%%%%%%%%%%%%%%%%% 
CHARM II
& LO 
& $ 0.39   ^{+0.07}_{-0.06}\; ^{+0.07}_{-0.07}\; $
& \cite{charm2}
\\ \hline  %%%%%%%%%%%%%%%%%%%%%%%%%%%%%%%% 
CCFR${}^*$
& NLO 
& $ 0.477 ^{+0.046}_{-0.044}\; ^{+0.023}_{-0.024}\; $
& \cite{bazarko}
\\ \hline  %%%%%%%%%%%%%%%%%%%%%%%%%%%%%%%%
CCFR${}^{\dagger}$
& NLO
& $ 0.468   ^{+0.061}_{-0.046}\; ^{+0.024}_{-0.025}\; $
& \cite{bazarko}
\\ \hline  %%%%%%%%%%%%%%%%%%%%%%%%%%%%%%%%
CCFR${}^{\dagger}$
& LO 
& $ 0.373  ^{+0.048}_{-0.041} \pm 0.018 $
& \cite{ccfrlo}
\\ \hline  %%%%%%%%%%%%%%%%%%%%%%%%%%%%%%%%
NOMAD
& LO 
& $ 0.48   ^{+0.09}_{-0.07}\; ^{+0.17}_{-0.12}\; $
& \cite{nomad}
\\ \hline  %%%%%%%%%%%%%%%%%%%%%%%%%%%%%%%%
NuTeV
& LO 
& $ 0.42 \pm {0.07} \pm {0.06} $
& \cite{nutev}
\\ \hline   %%%%%%%%%%%%%%%%%%%%%%%%%%%%%%%%
\hline   %%%%%%%%%%%%%%%%%%%%%%%%%%%%%%%%
\end{tabular}
 \caption{Next-to-leading-order and leading-order fit results. 
Errors are statistical and
 systematic.
This table is displayed to {\it estimate} 
the upper limits allowed by experiment;
a full comparison must take into account scheme and scale choices, and
the shape parameters. \protect\\
${}^*$Collins-Spiller Fragmentation. \protect\\
${}^\dagger$ Peterson  Fragmentation. 
} 
\end{center} 
\vspace{-2ex}
 \label{tab:kappa}
\end{table} 
}
\def\figone{
\begin{figure}[t] 
\begin{center}
\leavevmode
\vbox{
 \hbox{
 \epsfxsize=0.98\hsize \epsfbox{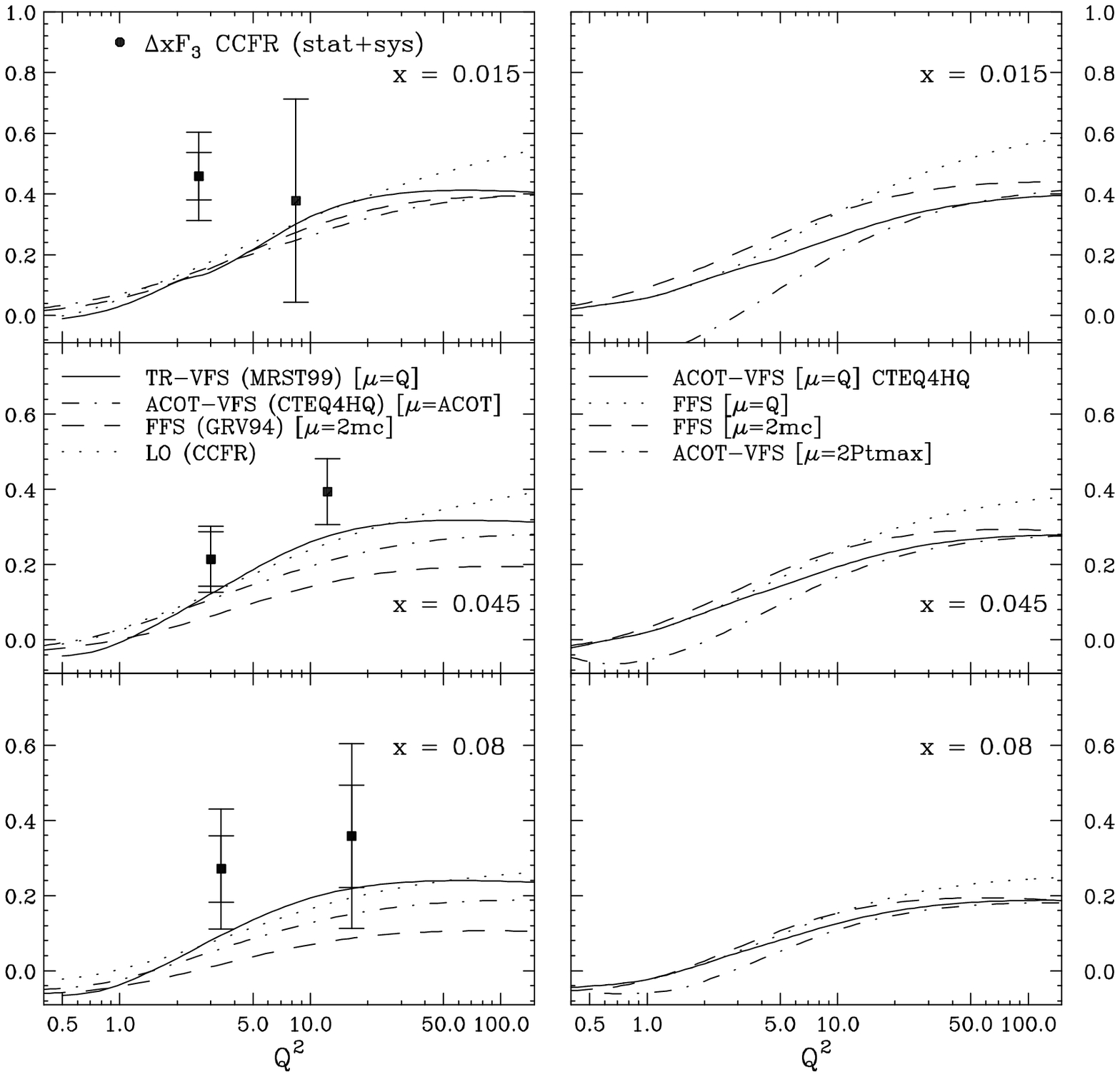} 
 }
}
\vskip -20pt
\caption{Comparison of $\Delta xF_3$ data from CCFR with various 
theoretical predictions. 
\label{fig:one} 
}
\vskip -20pt
\end{center}
\end{figure}
}
\def\figmc{
\begin{figure}[t] 
\begin{center}
\leavevmode
\vbox{
 \hbox{
 \epsfxsize=0.75\hsize \epsfbox{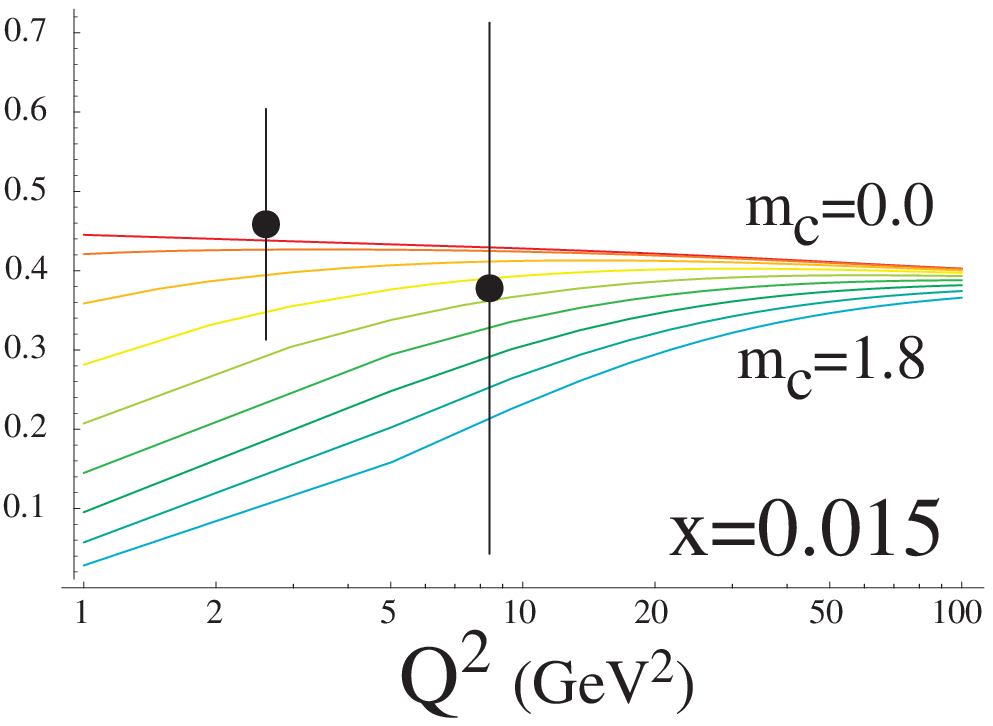} 
 }
 \hbox{
 \epsfxsize=0.75\hsize \epsfbox{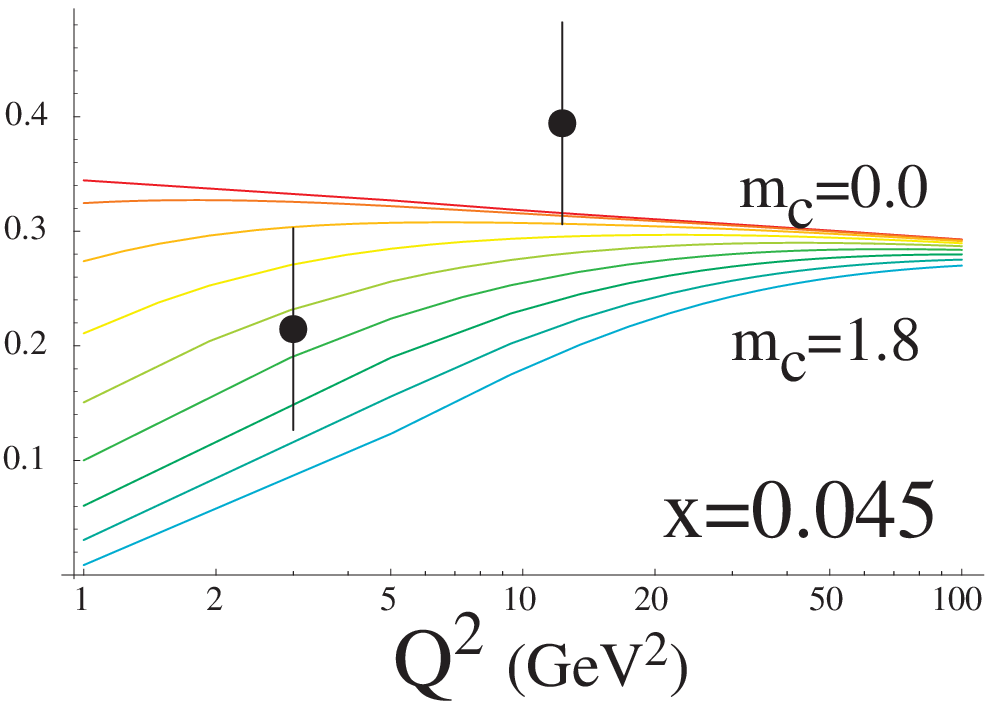}
 }
 \hbox{
 \epsfxsize=0.75\hsize \epsfbox{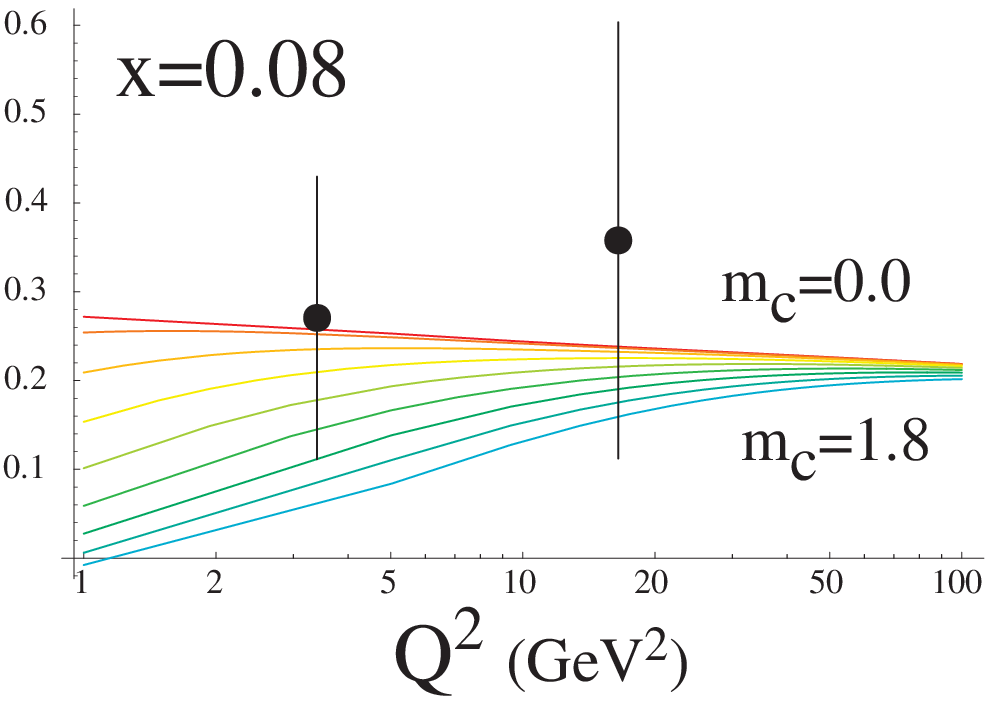}
 }
}
\vskip -20pt
\caption{
 Variation of the leading-order expression 
$\Delta xF_3 = 4 x \{s(\xi,\mu)- c(x,\mu)\}$ 
on the charm mass, $m_c$, plotted {\it vs.} $\mu$.
We take values of $m_c$ in the range $m_c=[0,1.8]$~GeV using steps of 
0.2~GeV for three choices of $x$. We use $\xi=x(1+m_c^2/Q^2)$, and 
$\mu^2=(Q^2+1.6^2 \, GeV^2)$.
\label{fig:mc} 
}
\vskip -20pt
\end{center}
\end{figure}
}
\def\figmu{
\begin{figure}[t] 
\begin{center}
\leavevmode
\vbox{
 \hbox{
 \epsfxsize=0.75\hsize \epsfbox{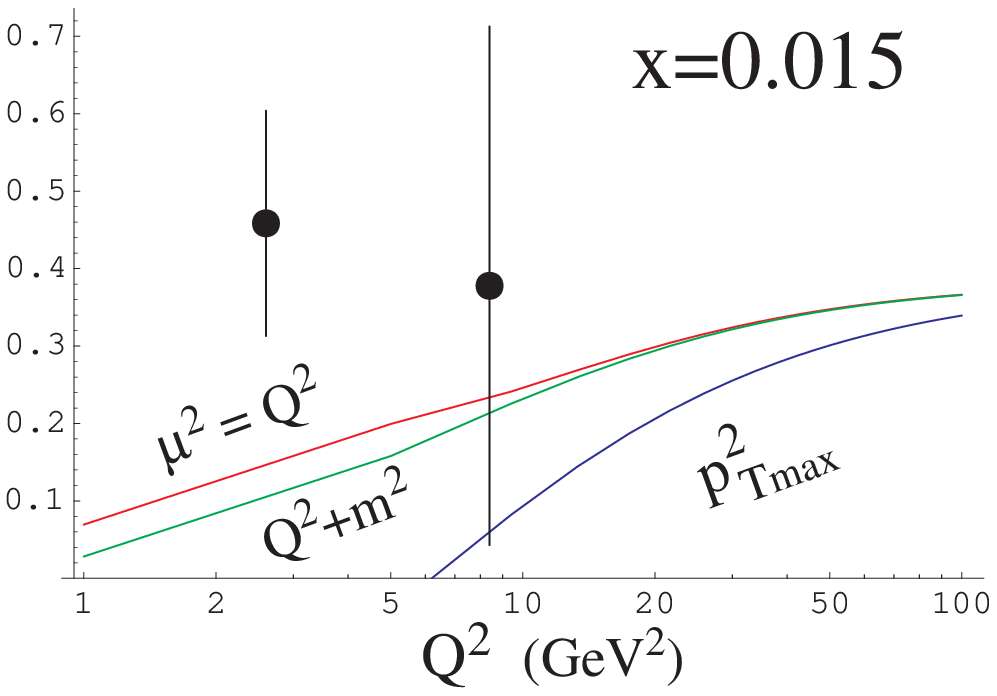} 
 }
 \hbox{
 \epsfxsize=0.75\hsize \epsfbox{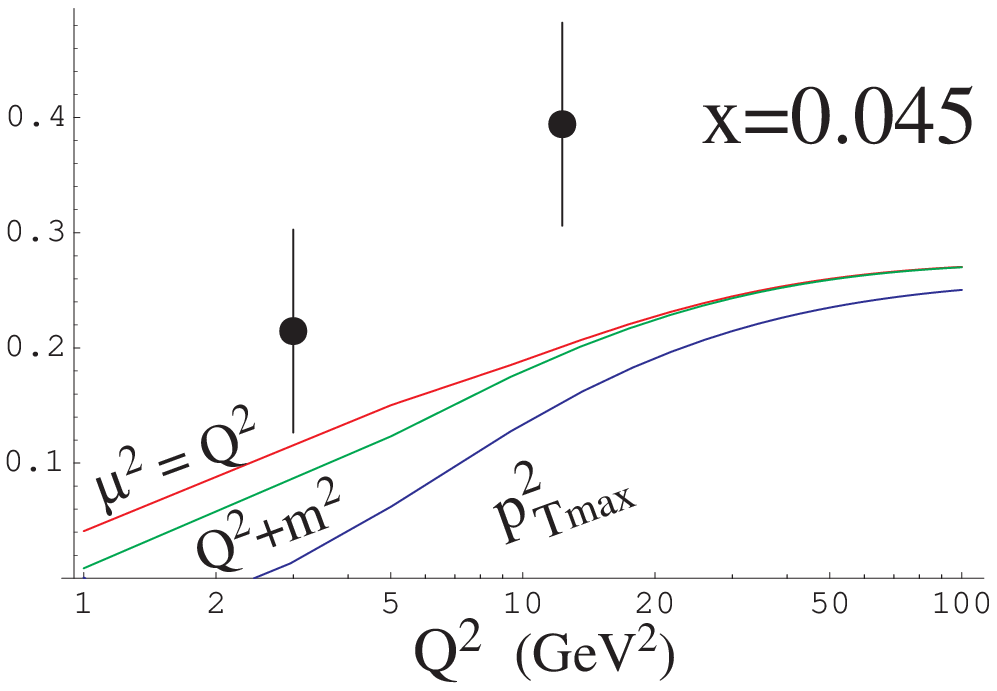}
 }
 \hbox{
 \epsfxsize=0.75\hsize \epsfbox{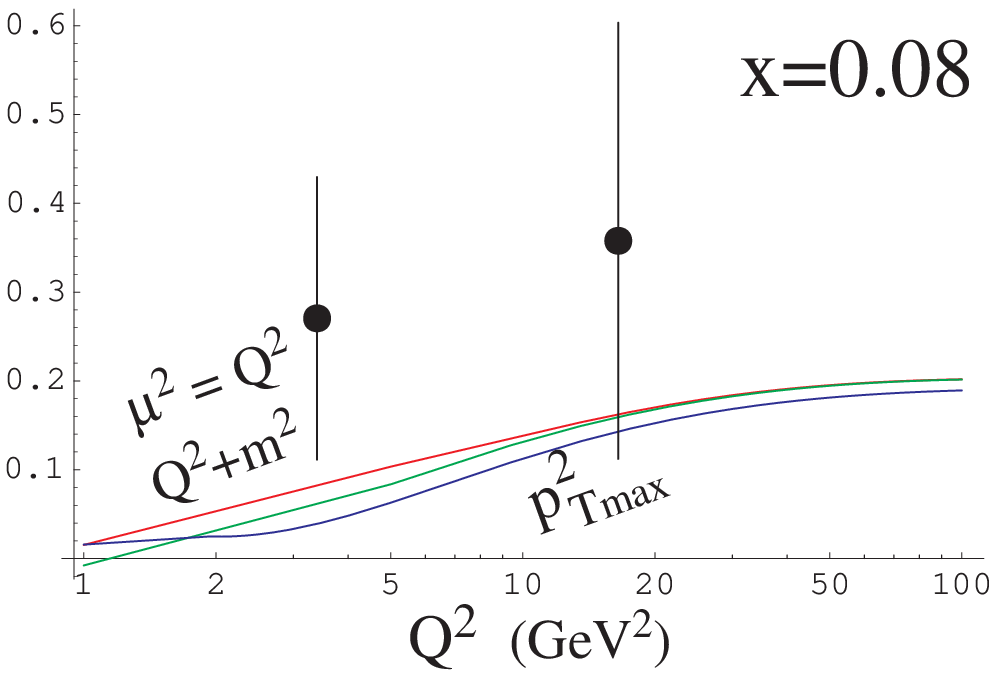}
 }
}
\vskip -20pt
\caption{
 Variation of the leading-order  
$\Delta xF_3$ on the renormalization scale, $\mu$. 
We have chosen $\mu^2 = \{Q^2, Q^2+m_c^2, \ptmax^2  \}$. 
\label{fig:mu} 
}
\vskip -20pt
\end{center}
\end{figure}
}
\def\figscheme{
\begin{figure}[t] 
\begin{center}
\leavevmode
\vbox{
 \hbox{
 \epsfxsize=0.85\hsize \epsfbox{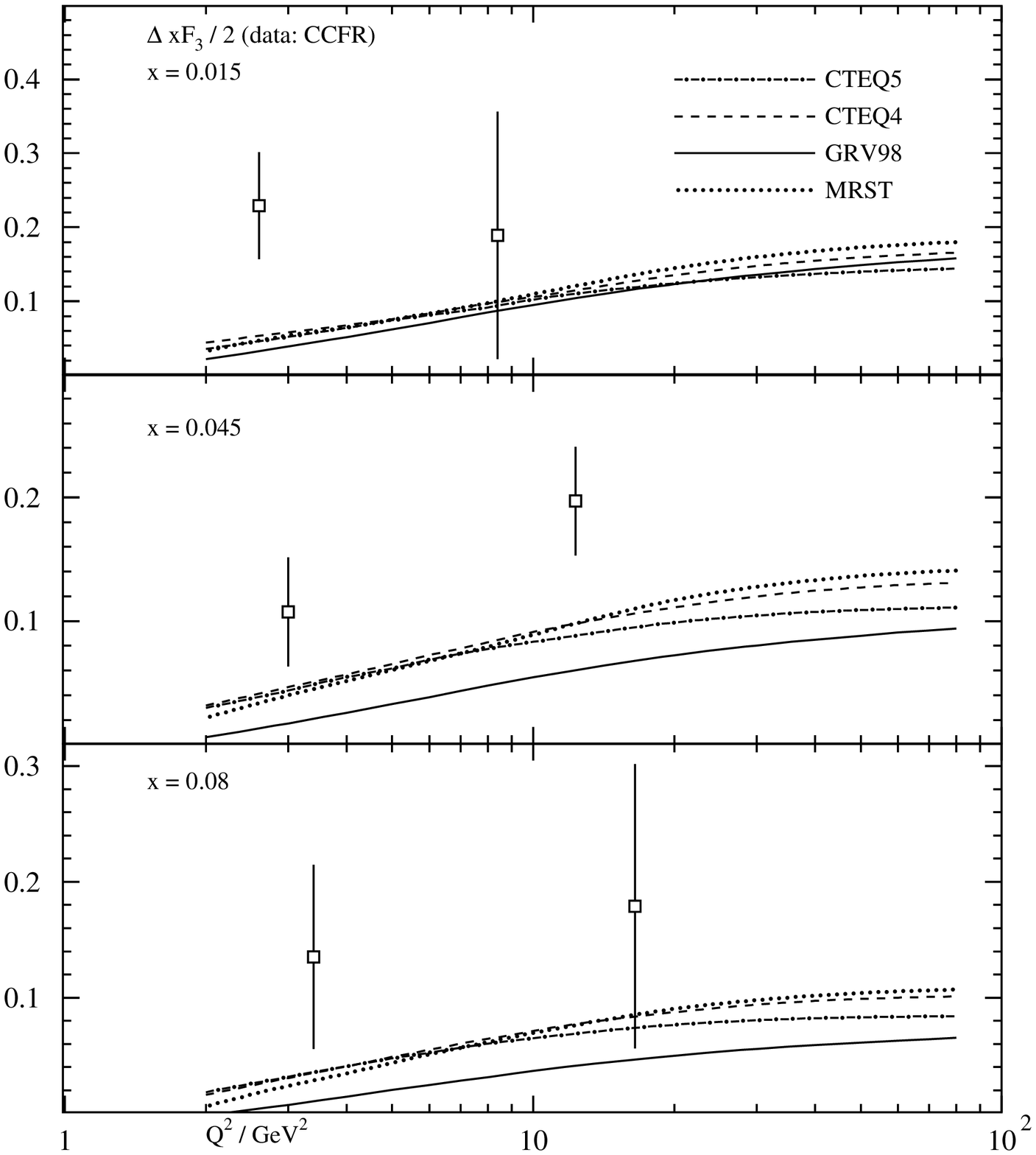}
 }
}
\vskip -20pt
\caption{
 Variation of $\Delta xF_3$ on the  renormalization scheme. 
All curves use NLO calculations, and appropriately matched PDF's.
We note that the theoretical predictions are consistent within the
theoretical uncertainty---as they should be. 
\label{fig:scheme} 
}
\vskip -20pt
\end{center}
\end{figure}
}
\def\fight{
\begin{figure}[t] 
\begin{center}
\leavevmode
\vbox{
 \hbox{ \epsfxsize=0.75\hsize \epsfbox{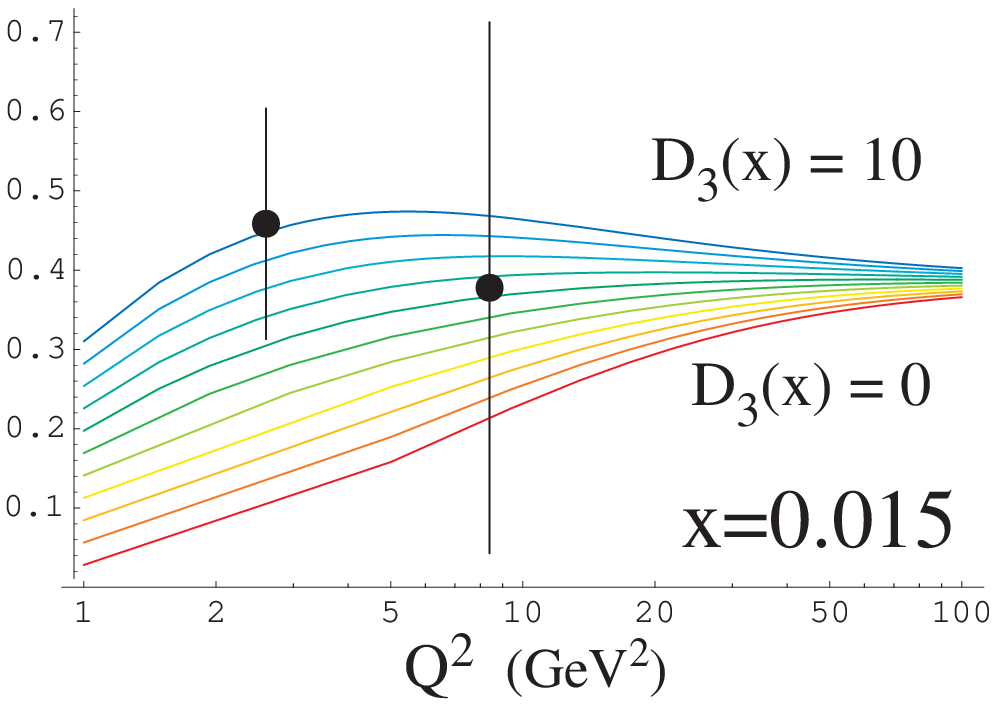} }
 \hbox{ \epsfxsize=0.75\hsize \epsfbox{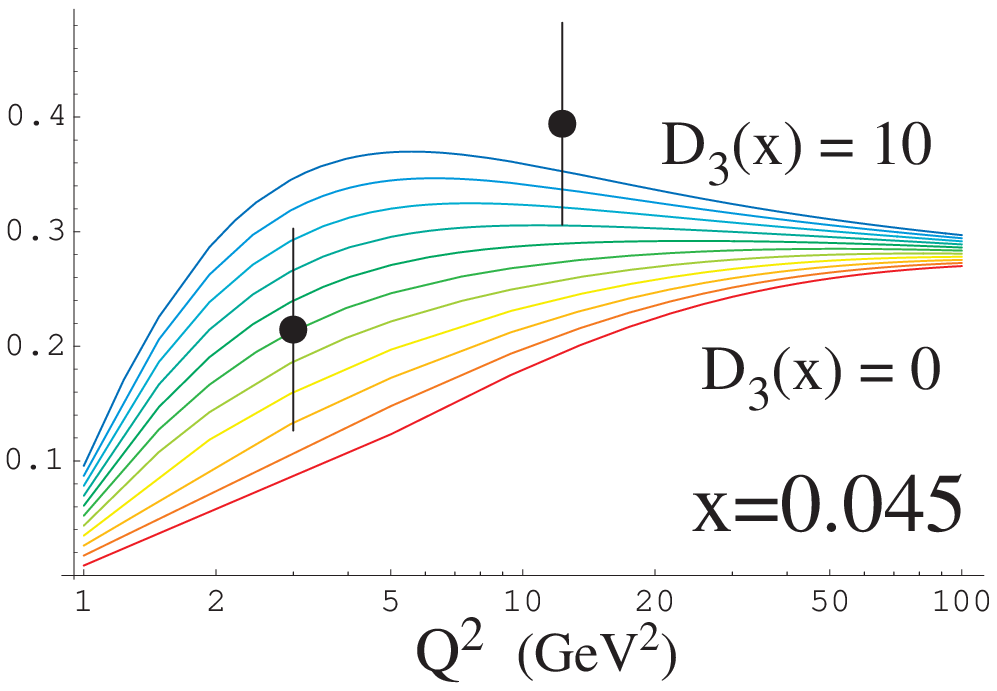} }
 \hbox{ \epsfxsize=0.75\hsize \epsfbox{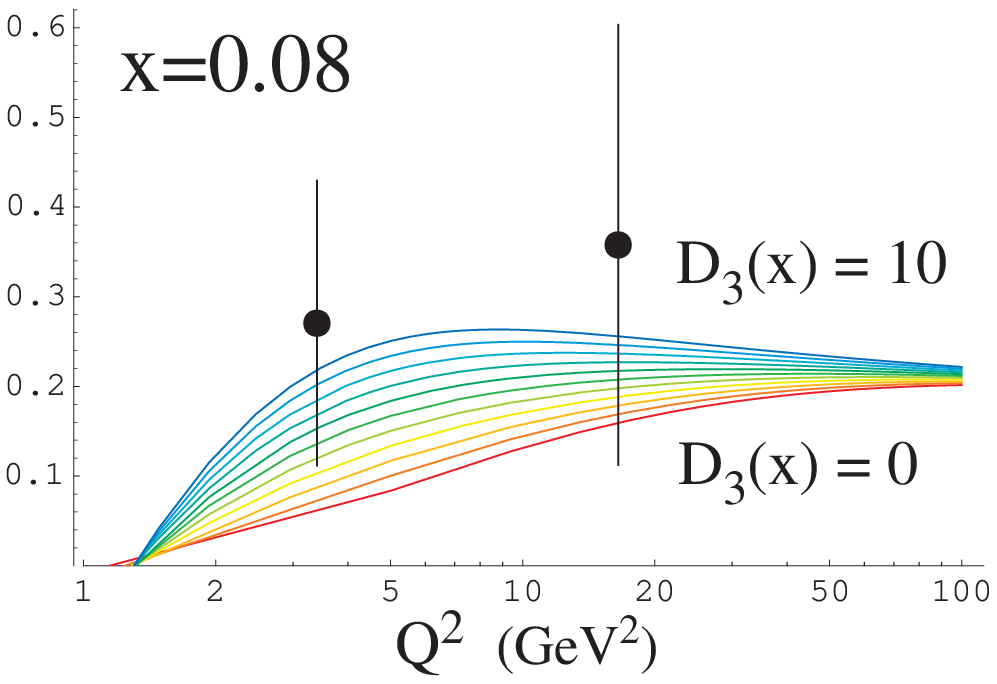} }
}
\vskip -20pt
\caption{
 Variation of the LO $\Delta xF_3$ on higher twist. 
The functional form is $(1+D_3(x)/Q^2)$ where
we take $D_3(x)$ in the range $D_3(x)=[0,10]~GeV^2$ in steps of 
$1~GeV^2$. 
\label{fig:ht} 
}
\vskip -20pt
\end{center}
\end{figure}
}
\def\figpdf{
\begin{figure}[t] 
\begin{center}
\leavevmode
\vbox{
 \hbox{
 \epsfxsize=0.85\hsize \epsfbox{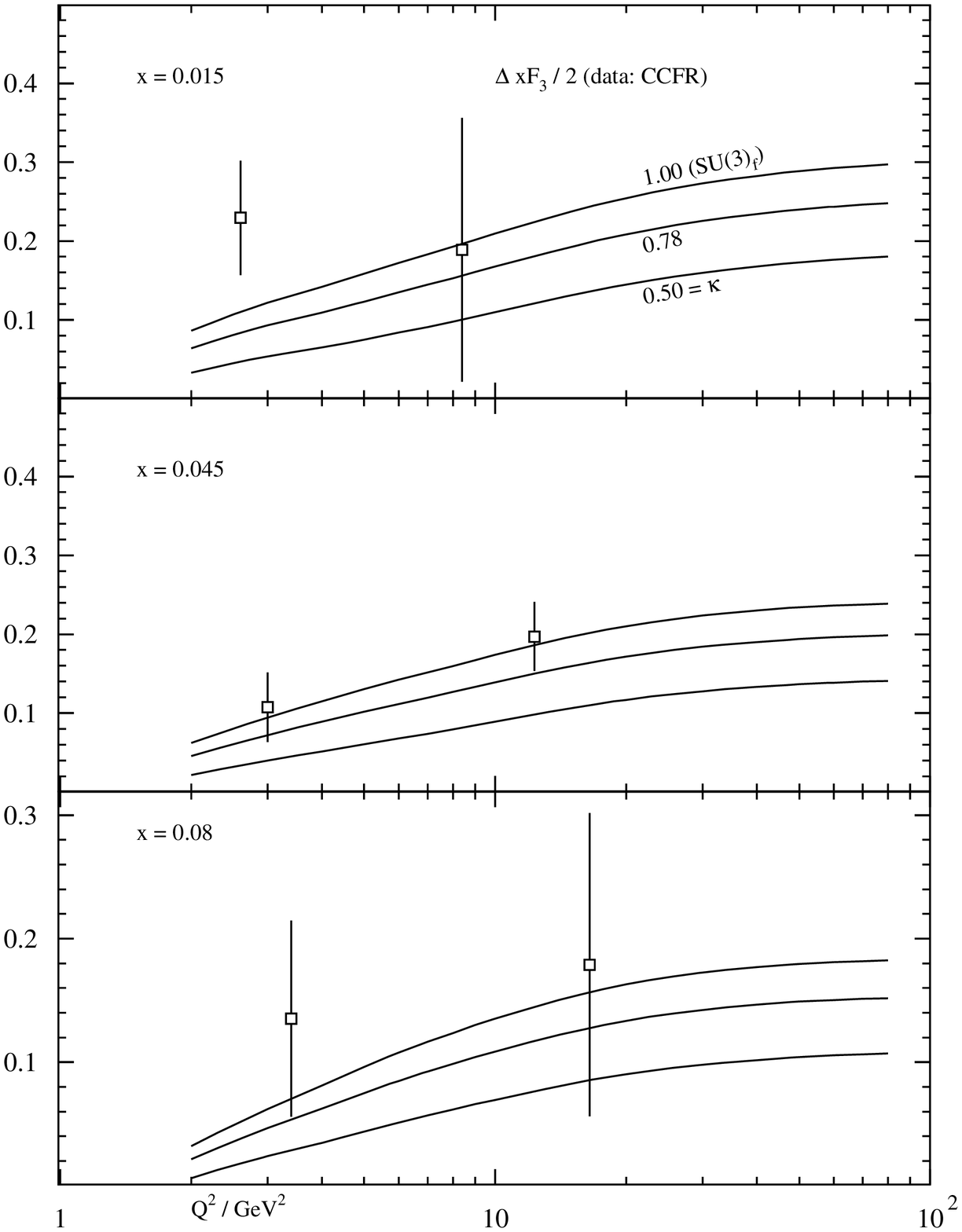} 
 }
}
\vskip -20pt
\caption{
 Variation of $\Delta xF_3$  
on the strange-quark PDF. 
The NLO calculation is used with three sets of PDF's. 
These PDF sets are re-fit based on the MRST set, and the value 
of $\kappa$ is constrained to be $\kappa=\{0.50, 0.78, 1.00 \}$.
\label{fig:pdf} 
}
\vskip -20pt
\end{center}
\end{figure}
}
\def\figisospin{
\begin{figure}[t] 
\begin{center}
\leavevmode
\vbox{
 \hbox{
 \epsfxsize=0.70\hsize \epsfbox{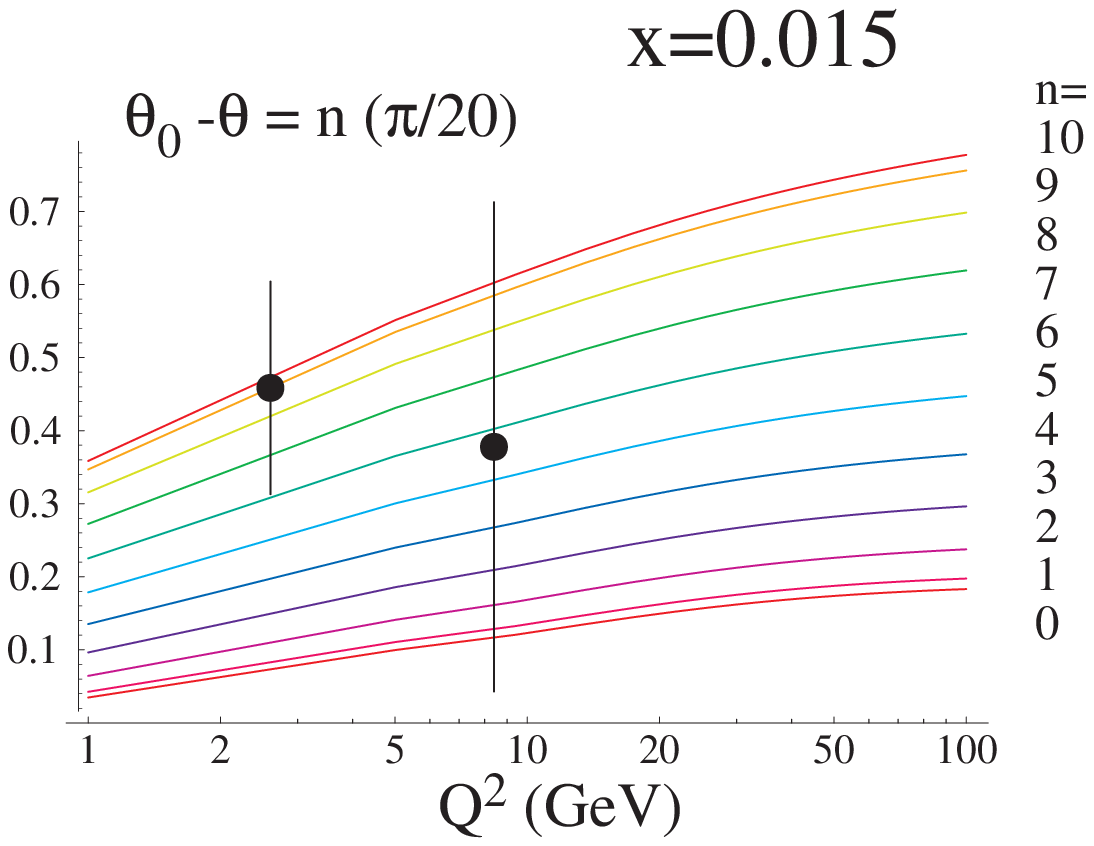} 
 }
 \hbox{
 \epsfxsize=0.70\hsize \epsfbox{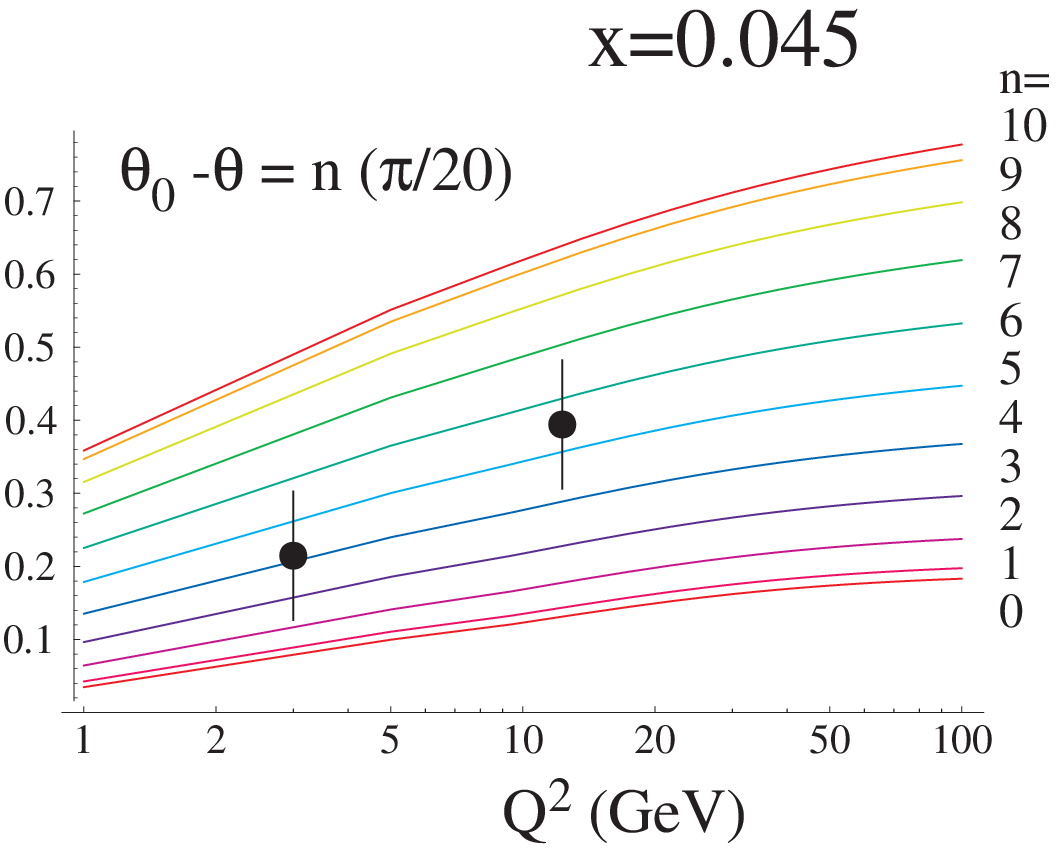}
 }
 \hbox{
 \epsfxsize=0.70\hsize \epsfbox{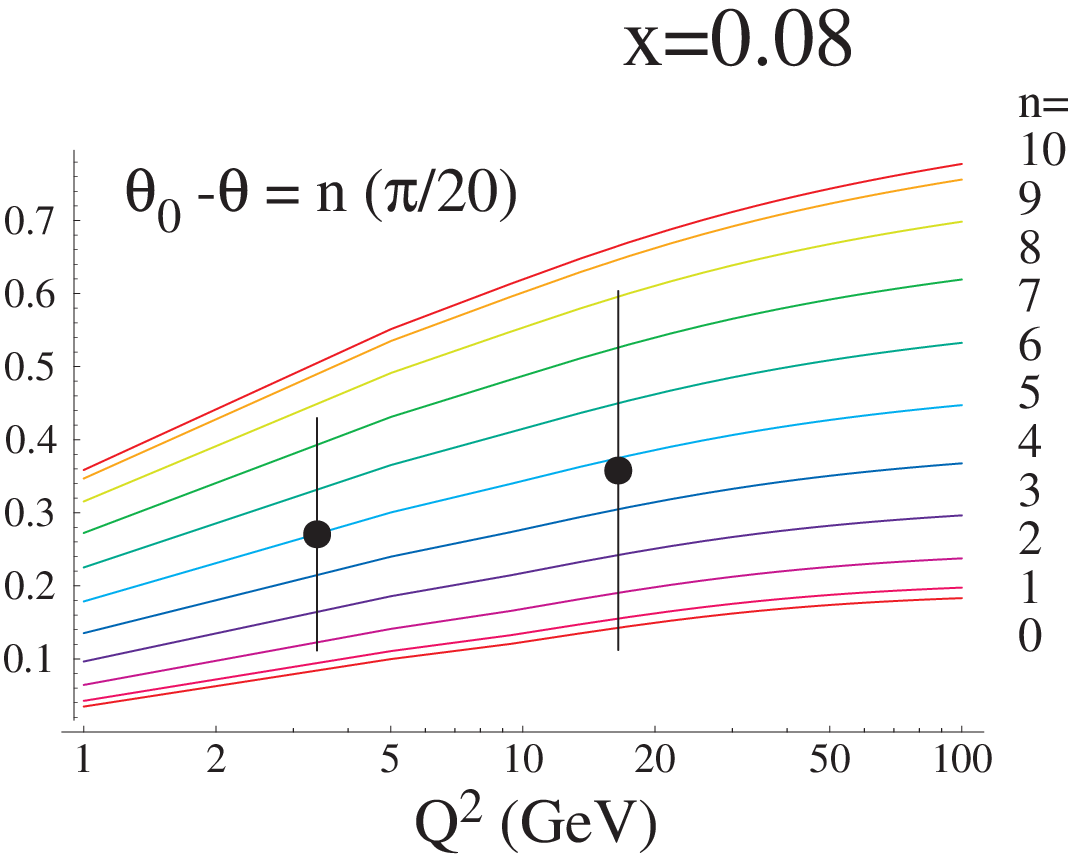}
 }
}
\vskip -20pt
\caption{
Impact of a one-parametric charge symmetry violation (CSV) toy model of 
\eq{toy1} and  \eq{toy2}
on the LO  $\Delta xF_3$ of \eq{lo2}.
$\theta$ is varied over the range  
$\theta_0 - \theta=[0,\pi/2]$  in steps of $(\pi/20)$.
\label{fig:isospin} 
}
\vskip -20pt
\end{center}
\end{figure}
}
\def\figiso{
\begin{figure}[t] 
\begin{center}
\leavevmode
\vbox{
 \hbox{
 \epsfxsize=0.85\hsize \epsfbox{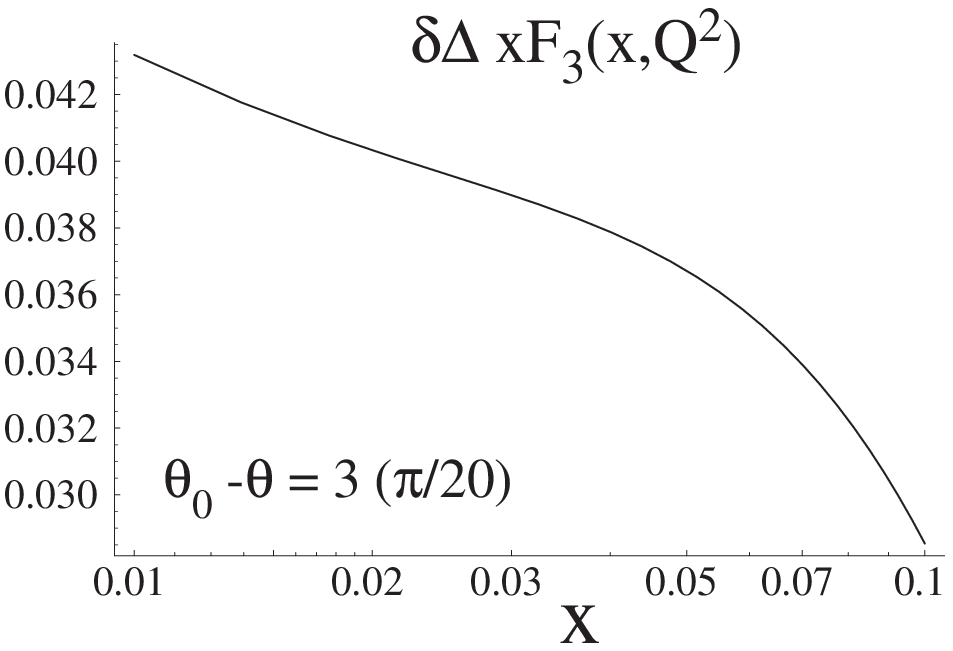} 
 }
}
\vskip -20pt
\caption{
$\delta\Delta xF_3(x,Q^2)$ with $Q^2=4\ {\rm{GeV}}^2$
as defined in \eq{Delta} and \eq{toy2} for 
$\theta_0 - \theta = 3 (\pi /20)$.
\label{fig:iso} 
}
\vskip -20pt
\end{center}
\end{figure}
}
\def\figspdf{
\begin{figure}[t] 
\begin{center}
\leavevmode
\vbox{
 \hbox{
 \epsfxsize=0.98\hsize \epsfbox{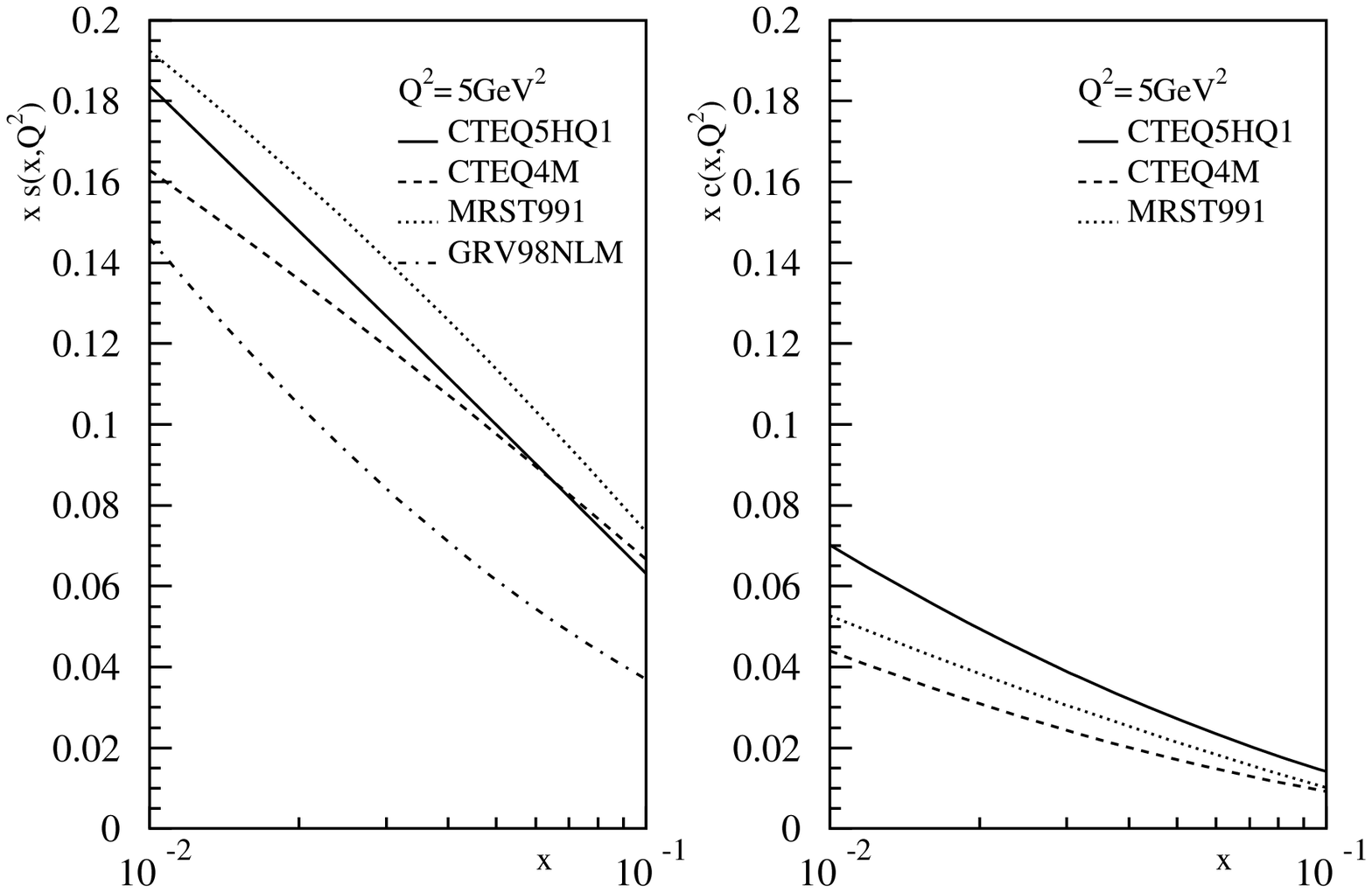} 
 }
}
\vskip -20pt
\caption{
a) We display several strange sea PDFs 
in the relevant $x$ range at $Q^2=5 \, GeV$. 
b) Same as above for the charm sea. 
\label{fig:spdf} 
}
\vskip -20pt
\end{center}
\end{figure}
}

\def\figxfnlo{
\begin{figure}[t] 
\begin{center}
\leavevmode
 \epsfxsize=0.85\hsize \epsfbox{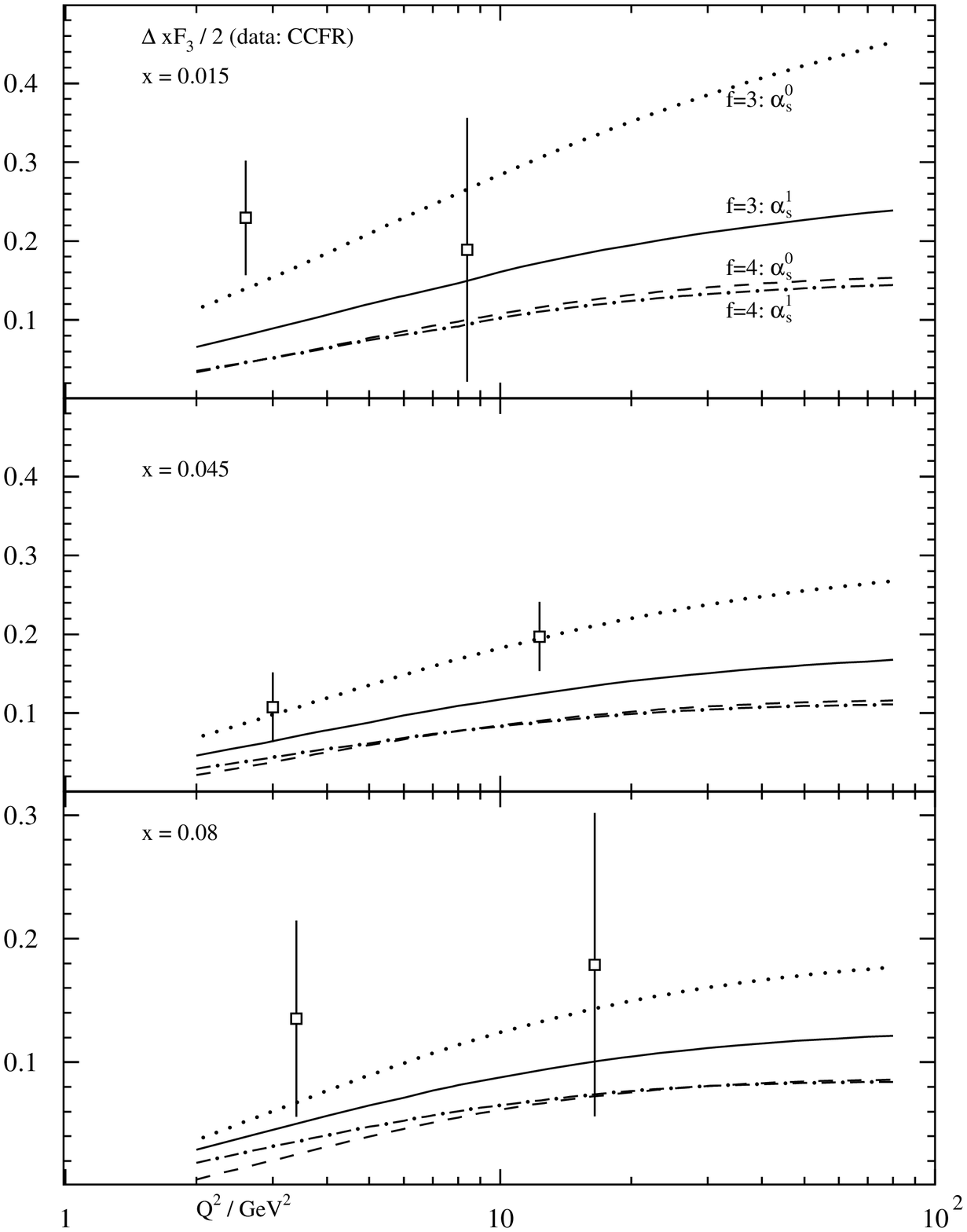} 
\vskip -20pt
\caption{
LO and NLO calculations for  $\Delta x F_3$ {\it vs.} $Q^2$
in 3 and 4 flavor schemes for  three values of $x$.
The dotted curve (f=3, $\alpha_s^0$) corresponds to the 3-flavor LO result.
The dashed curve (f=4, $\alpha_s^0$) corresponds to the 4-flavor LO result.
The solid and dot-dash curves correspond to the 3 and 4 flavor NLO QCD 
calculations, respectively. 
\label{fig:xfnlo} 
}
\vskip -20pt
\end{center}
\end{figure}
}
\def\figbazarko{
\begin{figure}[t] 
\begin{center}
\leavevmode
\vbox{
 \hbox{
 \epsfxsize=0.75\hsize \epsfbox{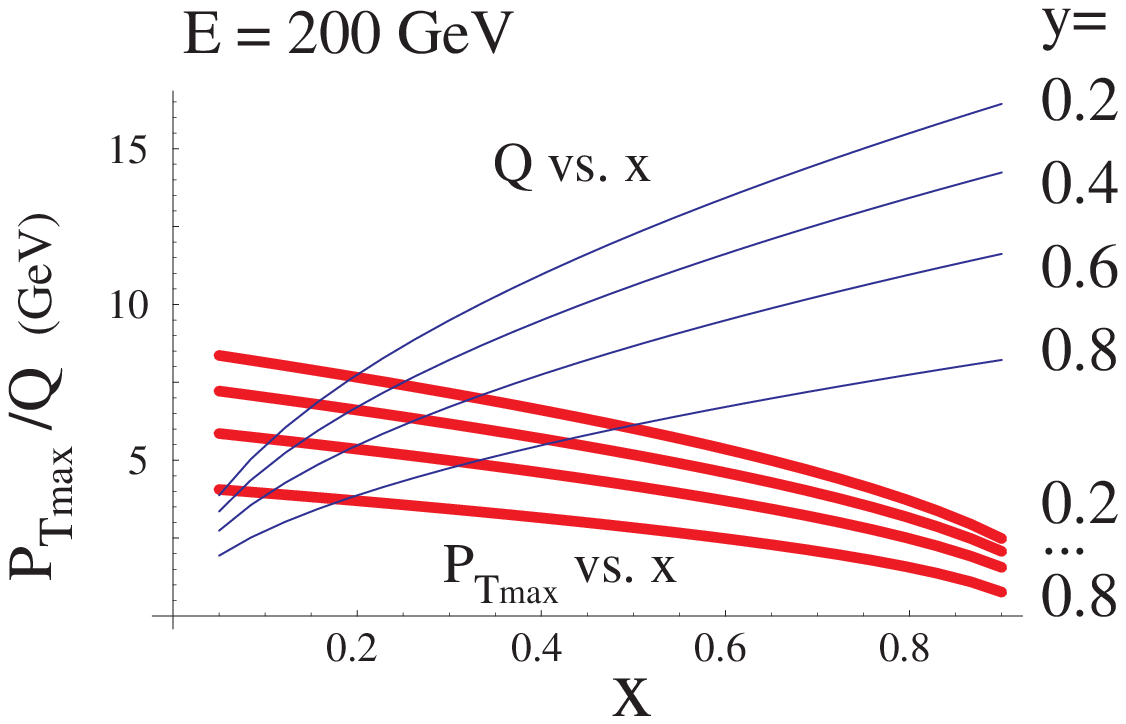} 
 }
 \hbox{
 \epsfxsize=0.75\hsize \epsfbox{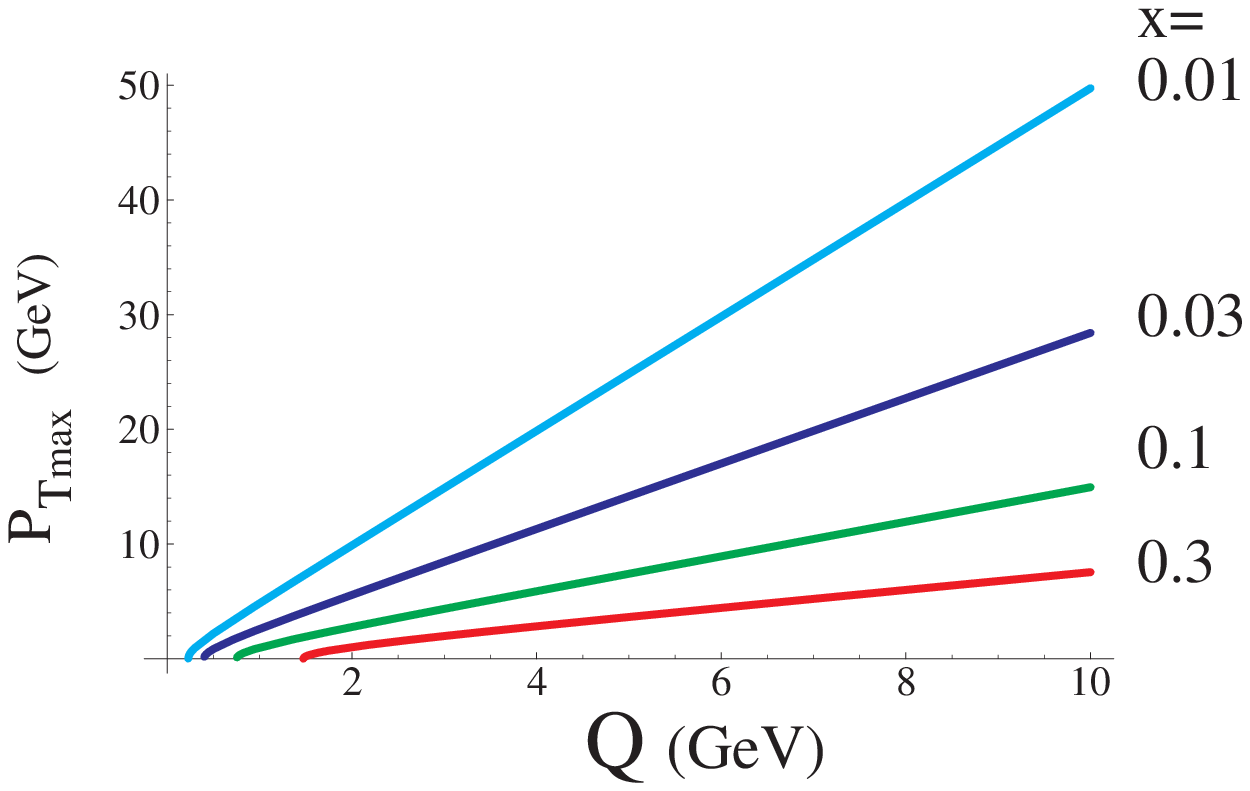}
 }
}
\vskip -20pt
\caption{
a) We display $P_T^{max}$ and $Q$ {\it vs.} $x$ for E=200~GeV and 
$y=\{ 0.2,0.4,0.6,0.8 \}$. 
b) $P_T^{max}$   {\it vs.} $Q$ for $x=\{ 0.3, 0.1, 0.03, 0.01 \}$.
\label{fig:bazarko} 
}
\vskip -20pt
\end{center}
\end{figure}
}
\begin{document}
%%%%%%%%%%%%%%%%%%%%%%%%%%%%%%%%%%%%%%%%%%%%%%%%%%%%%%%%%%%%%%%%%%%%%%%%%%%%%%

%%%%%%%%%%%%%%%%%%%%%%%%%%%%%%%%%%%%%%%%%%%%%%%%%%%%%%%%%%%%%%%%%%%%%%%%%%%%%%
\title{Predictions for Neutrino Structure Functions}
%%%%%%%%%%%%%%%%%%%%%%%%%%%%%%%%%%%%%%%%%%%%%%%%%%%%%%%%%%

\preprint{
 hep-ph/0101088  \\
January 2001\\
DO-TH 2000/19\\
MSUHEP-01205\\
Cavendish-HEP-2001/01
}

\def\addr#1{\address{#1 \vskip -8pt}}

\author{
S.~Kretzer\rlap,\addr{Univ. Dortmund, Dept. of  Physics, 
      D-44227 Dortmund, Germany
}${}^{,}$\addr{Michigan State University, Dept. of Physics and Astronomy,
East Lansing, MI 48824-1116}
F.I.~Olness\rlap,\addr{Southern Methodist University, Department of Physics, 
      Dallas, TX 75275-0175}
R.J.~Scalise\rlap,$^{\rm c}$
R.S.~Thorne\rlap,\addr{Cavendish Laboratory, 
University of Cambridge, Madingley Road, Cambridge, CB3 0HE}\thanks{
Royal Society University Research Fellow}
U.K.~Yang\addr{University of Chicago, Enrico Fermi Institute, 
      Chicago, IL 60637-1434
}${}^{,}$\addr{Univ. of Rochester, Rochester, NY 14627}
}

%%%%%%%%%%%%%%%%%%%%%%%%%%%%%%%%%%%%%%%%%%%%%%%%%%%

\begin{abstract}

 The first measurements of $\Delta xF_3$ are higher
than current theoretical predictions.
 We investigate the sensitivity of these theoretical predictions
upon a variety of factors including: 
renormalization scheme and scale, 
quark mass effects, higher twist, isospin violation, and PDF uncertainties. 

\end{abstract}

\maketitle

%%%%%%%%%%%%%%%%%%%%%%%%%%%%%%%%%%%%%%%%%%%%%%%%%%%%%%%%%%%%%%%%%%%%%%%%%%%%%%
\section{Introduction} 
%%%%%%%%%%%%%%%%%%%%%%%%%%%%%%%%%%%%%%%%%%%%%%%%%%%%%%%%%%%%%%%%%%%%%%%%%%%%%%

%%%%%%%%%%%%%%%%%%%%%%%%%%%%%%%%%%%%%%%%%%%%%%%%%%%%%%%%%%%%%%%%%%%%%%%%%%%%%%
%\subsection{Overview} 
%%%%%%%%%%%%%%%%%%%%%%%%%%%%%%%%%%%%%%%%%%%%%%%%%%%%%%%%%%%%%%%%%%%%%%%%%%%%%%

Deep inelastic lepton-nucleon scattering experiments have
provided precision information about the quark distributions in the nucleon.
 However, there has been a long-standing discrepancy between the $F_2$ structure 
functions extracted from neutrino and muon experiments in the 
small $x$ range. 
 Recently, a new analysis of  differential cross sections
and structure functions from CCFR $\nu_\mu$-Fe and $\nub_\mu$-Fe data
was presented; in this study,  the neutrino-muon difference is resolved by
extracting the $\nu_\mu$ structure functions in a physics
model independent way.\cite{yang}

In previous analyses  
of $\nu_\mu$ data,\cite{ccfrsf}
structure functions were extracted
by applying a slow rescaling correction to correct for the charm
mass suppression in the final state. In addition, 
the $\Delta xF_3$ term (used as input in the extraction)
was calculated from a leading order charm production model. These
resulted in physics model dependent (PMD) structure functions. 
 In the
new analysis,\cite{yang} slow rescaling corrections are not applied,
 and $\Delta xF_3$ and $F_2$
were extracted from two parameter fits to the data.

The extracted physics model independent (PMI) values for  $F_2^{\nu}$
are then compared  with $F_2^{\mu}$
within the framework of NLO models for massive charm production;
these are found to be in agreement, thus resolving the long-standing 
discrepancy between the two
sets of data. 
 However, the first measurements of $\Delta xF_3$ are higher
than current theoretical predictions.
 The objective of this paper is to investigate the sensitivity of  
$\Delta xF_3$ upon a variety of factors including renormalization scheme 
and scale, 
quark mass effects, higher twist, isospin violation, and PDF uncertainties.

%%%%%%%%%%%%%%%%%%%%%%%%%%%%%%%%%%%%%%%%%%%%%%%%%%%%%%%%%%%%%%%%%%%%%%%%%%%%%%
 \figone

In \fig{one} we have plotted the quantity $\Delta xF_3$ 
for an isoscalar target computed to 
order $\alpha_s^1$. 
 For comparison, we also display data from the 
CCFR analysis.\cite{yang} While there is much freedom in the
theoretical calculation, the difference between these calculations
and the data at low $Q$ values warrants further investigation.
 We will discuss  and compare the different theoretical 
calculations, and examine the inherent uncertainty in each 
with respect to different input parameters. We will also examine 
the experimental input, and assess uncertainties in this data.

%%%%%%%%%%%%%%%%%%%%%%%%%%%%%%%%%%%%%%%%%%%%%%%%%%%%%%%%%%%%%%%%%%%%%%%%%%%%%%
\subsection{Measurement of $\Delta xF_3$} 
%%%%%%%%%%%%%%%%%%%%%%%%%%%%%%%%%%%%%%%%%%%%%%%%%%%%%%%%%%%%%%%%%%%%%%%%%%%%%%

The structure functions are defined in 
terms of the neutrino-nucleon 
cross section via: 
\begin{eqnarray}
\frac{d^2 \sigma^{\nu,\bar{\nu}}}{dx \, dy}
=
\frac{\scriptstyle G_F^2 M E_{\nu}}{\pi}
\left[ 
F_2 (1-y) + x F_1 y^2 \pm x F_3 y (1-\frac{y}{2}) 
\right]
\nonumber \\
\label{eq:}
\end{eqnarray}
where $G_{F}$ is the Fermi weak coupling constant, 
$M$ is the nucleon mass, 
$E_{\nu}$ is the incident energy, 
$y=E_h/E_{\nu}$ is the fractional energy transfer, 
 and
$E_h$ is the final state hadronic energy. 

The sum of $\nu_\mu$ and $\nub_\mu$
 differential cross sections 
for charged current interactions 
on an isoscalar target is then:
\begin{eqnarray}
F(\epsilon) 
&\equiv&
\left[ 
  \frac{d^2\sigma^{\nu }}{dxdy}
 +\frac{d^2\sigma^{\overline \nu}}{dxdy} 
\right] 
 \frac{(1-\epsilon)\pi}{y^2G_F^2ME_\nu}
\nonumber\\
&=& 
2xF_1 [ 1+\epsilon R ] + \frac{y(1-y/2)}{1+(1-y)^2} \Delta xF_3 
\label{eq:feps}
\end{eqnarray}
where $\epsilon\simeq2(1-y)/[1+(1-y)^2]$ is the polarization of 
the virtual $W$ boson.
 In the above equation, we have used the relation: 
\begin{equation}
2xF_1(x,Q^2)
=
F_2(x,Q^2) \  \frac{1+4M^2x^2/Q^2}{1+R(x,Q^2)}
\label{eq:f12r}
\end{equation}
where  $R=\sigma_{L}/\sigma _{T}$ 
is the ratio of the cross-sections of
longitudinally- to transversely-polarized $W$-bosons,
$Q^2$ is the square of the four-momentum transfer to the nucleon,
and 
$x=Q^2/2ME_h$ is the Bjorken scaling variable.
For $x<0.1$,  $R$ in neutrino scattering is expected to be somewhat larger
than $R$ for muon scattering because 
of the production of massive charm quarks
in the final state for the charged current neutrino production.

Using \eq{feps} and \eq{f12r}, $F_2$ and $\Delta xF_3$ can be extracted 
separately.
 Because of the positive correlation between
$R$ and $\Delta xF_3$, the extracted values of $F_2$ are rather insensitive
to the input $R$. If a large input $R$ is used, a larger value of $xF_3$
is extracted from the $y$ distribution, thus yielding the
same value of $F_2$. 
 In contrast, the extracted values of $\Delta xF_3$
are sensitive to the assumed value of $R$, which yields 
a larger systematic error, shown on the data. For $R$ the QCD-inspired fit 
in \cite{whitlow} is used, but corrected for 
charged current neutrino scattering using a
leading order slow rescaling model. This gives precisely the same type of 
correction as a full NLO calculation including the massive charm quark, as
shown in Ref.~\cite{bslt}, 
but leads to a somewhat higher normalization than the
perturbative correction. It is arguable which prescription for $R$ leads to 
the better fit to 
existing data, but the difference between the two leads to an uncertainty 
in $\Delta xF_3$ of
the order of the systematic error shown. Clearly a further reduction in 
the assumed value of R (even down to zero), as suggested by models including 
$\ln(1/x)$ resummations, would still leave a discrepancy 
between the lowest $Q^2$ data points and the theoretical predictions.

%%%%%%%%%%%%%%%%%%%%%%%%%%%%%%%%%%%%%%%%%%%%%%%%%%%%%%%%%%%%%%%%%%%%%%%%%%%%%%
\subsection{Quark Parton Model Relations} 
%%%%%%%%%%%%%%%%%%%%%%%%%%%%%%%%%%%%%%%%%%%%%%%%%%%%%%%%%%%%%%%%%%%%%%%%%%%%%%

Now that we have outlined the experimental method used in the extraction of 
the $F$ structure functions, 
it is instructive to recall the simple leading-order correspondence 
between the $F$'s and the PDF's:\addtocounter{footnote}{-1}\footnote{%
To exhibit the basic structure, 
the above is taken in the limit of 4 quarks, a symmetric sea, and a 
vanishing Cabibbo angle.
 Of course, the actual analysis takes into account the full
structure.\cite{yang}
}
\begin{eqnarray}
F_2^{(\nu,\bar{\nu}) N} &\simeq&
  x 
\left\{ 
u + \bar{u} + d + \bar{d} + 2s + 2c
\right\} 
\nonumber \\
x F_3^{(\nu,\bar{\nu}) N} &\simeq&
 x 
\left\{ 
u - \bar{u} + d - \bar{d} \pm 2s \mp 2c
\right\} 
\label{eq:lo1}
\end{eqnarray}
 Therefore, the combination $\Delta xF_3$ yields:
\begin{equation}
\Delta xF_3 = x F_3^{\nu N} - x F_3^{\bar{\nu} N} 
\simeq
4 x \{ s - c \}
\quad .
\label{eq:lo2}
\end{equation}
Note that since this quantity involves the parity violating structure
function $F_3$, this measurement has no analogue in the neutral
current photon-exchange process.  Also note that since, at
leading-order, $\Delta xF_3$ is directly sensitive to the strange and
charm distributions, this observable can be used to probe the heavy
quark PDF's, and to understand heavy quark (charm) production.  We
discuss these possibilities further in the following subsection.

%%%%%%%%%%%%%%%%%%%%%%%%%%%%%%%%%%%%%%%%%%%%%%%%%%%%%%%%%%%%%%%%%%%%%%%%%%%%%%
\subsection{Implications for PDF's} 
%%%%%%%%%%%%%%%%%%%%%%%%%%%%%%%%%%%%%%%%%%%%%%%%%%%%%%%%%%%%%%%%%%%%%%%%%%%%%%

We have illustrated in \eq{lo2} how $\Delta xF_3$ is closely tied to
the heavy quark PDF's. The questions is: 
given the present knowledge
base, should we use $\Delta xF_3$ to determine the heavy quark PDF'S,
or vice verse.  To answer this question, we briefly review present
measurements of heavy quark PDF's, and assess their uncertainty.

%%%%%%%%%%%%%%%%%%%%%%%%%%%%%%%%%%%%%%%%%%%%%%%%%%%%%%%%%%%%%%%%%%%%%%%%%%%%%%
\subsubsection{Tevatron $W+Q$ Production} 
%%%%%%%%%%%%%%%%%%%%%%%%%%%%%%%%%%%%%%%%%%%%%%%%%%%%%%%%%%%%%%%%%%%%%%%%%%%%%%

The precise measurement of $W$ plus heavy quark $(W+Q)$ events
provides important information on heavy quark PDF's; additionally,
such signals are a background for Higgs and squark
searches.\cite{gkl,run2}

Unfortunately, a primary uncertainty for $W+Q$ production 
comes from the heavy quark PDF's. 
 Given that $\Delta xF_3$ is sensitive to these heavy quark PDF's, 
we see at least two scenarios. 
  One possibility is that new analysis of present data will resolve this
situation
prior to Run~II, and provide precise distributions as an input to the
Tevatron data analysis. 
If the situation remains unresolved, then new data
from Run~II may help to finally solve this puzzle. 
In the far future, a neutrino experiment from the neutrino factory 
at a linear collider would be an ideal 
tool to measure any neutrino structure function.\cite{bhm}

%%%%%%%%%%%%%%%%%%%%%%%%%%%%%%%%%%%%%%%%%%%%%%%%%%%%%%%%%%%%%%%%%%%%%%%%%%%%%%
\subsubsection{DIS Di-muon production}
%%%%%%%%%%%%%%%%%%%%%%%%%%%%%%%%%%%%%%%%%%%%%%%%%%%%%%%%%%%%%%%%%%%%%%%%%%%%%%

The strange distribution is directly measured by dimuon production in 
neutrino-nucleon 
scattering.\footnote{%
Presently, there are a number of LO analyses, 
and one NLO ACOT analysis.\cite{cdhsw,ccfrlo,nutev,charm2}.
Results of a recent LO analysis by NOMAD \cite{nomad} are in line 
with these experiments.}
The basic sub-process is $\nu s \to \mu^- c X$ with a subsequent charm
decay $c \to \mu^+ X'$. 

DIS dimuon measurements have safely established the breaking of the
$SU(3)$ flavor symmetry,
\begin{equation}
\kappa \equiv 
\left.
\frac{\int dx\ x\ s(x)}{ \int dx\ x\ [{\bar u}(x)+{\bar d}(x)]/2}
\right|_{Q^2={{\cal{O}}(10{\rm GeV^2})}}
\lesssim \frac{1}{2}
\label{eq:kappa}
\end{equation} 
in the nucleon sea.
 Still, there remain large uncertainties for the $s$-quark distribution 
in the  kinematic regions relevant for $\Delta xF_3$,
even 
though the $(x,Q)$-range of the CCFR dimuon measurements \cite{ccfrlo,bazarko}
is comparable.
 CCFR recorded 5044 $\nu_\mu$ and 1062 $\nub_\mu$ \ $\mu^\mp \mu^\pm$ events
with $P_{\mu_1}\geq 9\, GeV$,   $P_{\mu_2}\geq 5\, GeV$,  
 $30 \geq E_{\nu} \geq 600\, GeV$,  and $\langle Q^2 \rangle = 22.2 \, GeV^2$.
The more recent NuTeV experiment recorded 
a similar sample of events, and these  
are presently being analysed through a
MC simulation based on NLO quark- and gluon-initiated 
corrections at differential level\cite{nutev,gkr2,koinprep}. 
A complete NLO analysis of this data together with a global analysis 
may help to further constraint the strange quark distribution.

At present, PDF sets take strangeness suppresssion into account
by imposing the constraint in \eq{kappa} as 
$s(x)=\kappa [{\bar u}(x) + {\bar d}(x)]/2$ 
at the PDF input scale $Q_0 \simeq 1\ {\rm GeV}$,
\cite{MRST,cteq5} or by evolving $s(x)$ from a 
vanishing input at a lower scale.\cite{grv98}
The residual uncertainty can be large, as can be seen from the collection
of strange seas in \fig{spdf}.

%%%%%%%%%%%%%%%%%%%%%%%%%%%%%%%%%%%%%%%%%%%%%%%%%%%%%%%%%%%%%%%%%%%%%%%%%%%%%%
\figspdf
  
In the following, we will take the uncertainty in the strange PDF into account
by relaxing the experimental constraint in \eq{kappa} as implemented in
the MRST partons.\cite{MRST} This will be discussed in 
\sec{squark}.

%%%%%%%%%%%%%%%%%%%%%%%%%%%%%%%%%%%%%%%%%%%%%%%%%%%%%%%%%%%%%%%%%%%%%%%%%%%%%%
\subsubsection{Charged and Neutral Current DIS} 
%%%%%%%%%%%%%%%%%%%%%%%%%%%%%%%%%%%%%%%%%%%%%%%%%%%%%%%%%%%%%%%%%%%%%%%%%%%%%%

The strange distribution can also be extracted indirectly using a
combination 
of charged  ($F_2^{\nu}$) and neutral ($F_2^{\mu}$) current structure functions; 
however, the systematic uncertainties
involved
in this procedure make an accurate determination
difficult.\cite{yang}
The basic idea is to use the (leading-order) relation
 \begin{equation}
\frac{F_2^{\mu}}{F_2^{\nu}} 
\simeq
\frac{5}{18}
\left\{ 
1 - \frac{3}{5} \, \frac{(s+\bar{s})-(c+\bar{c}) + ...}{q+\bar{q}}
\right\}  
\end{equation}
to extract the strange distribution. 
 Here,  $q+\bar{q}$ represents a sum over all quark flavors. 
This method is complicated
by a number of issues including the $xF_3$ component which can 
play a crucial role in the small-$x$ region---precisely
the region where we observe the discrepancy.
 From the corresponding relation: 
 \begin{equation}
\frac{5}{18} F_2^{\nu} - F_2^{\mu} 
\simeq
\frac{1}{12} \Delta x F_3 
\end{equation}
we see that these problems are not independent;
 however,  this information, together with the exclusive dimuon events,
may provide a more precise determination of the strange quark sea,
and help to resolve our puzzle. 

Prior to the DIS dimuon data,
the 1992 CTEQ1 analysis \cite{cteq1} found that a combination of
NC structure functions from NMC\cite{nmc} and 
the physics-model-dependent charged current structure functions from CCFR
\cite{ccfrlo} seem compatible with approximate $SU(3)$-symmetry, 
{\it i.e.}, $\kappa \sim 1$ in \eq{kappa}.   
Recent dimuon measurements now exclude an $SU(3)$-symmetry $\kappa =1$. 

We shall explore the effect of $\kappa$ on $\Delta x F_3$ in 
\sec{squark}.

%%%%%%%%%%%%%%%%%%%%%%%%%%%%%%%%%%%%%%%%%%%%%%%%%%%%%%%%%%%%%%%%%%%%%%%%%%%%%%
%%%%%%%%%%%%%%%%%%%%%%%%%%%%%%%%%%%%%%%%%%%%%%%%%%%%%%%%%%%%%%%%%%%%%%%%%%%%%%
\section{Dependence of $\Delta xF_3$ on input parameters}
%%%%%%%%%%%%%%%%%%%%%%%%%%%%%%%%%%%%%%%%%%%%%%%%%%%%%%%%%%%%%%%%%%%%%%%%%%%%%%
%%%%%%%%%%%%%%%%%%%%%%%%%%%%%%%%%%%%%%%%%%%%%%%%%%%%%%%%%%%%%%%%%%%%%%%%%%%%%%

We now systematically investigate the 
sensitivity of the theoretical predictions 
of $\Delta xF_3$ upon a variety of 
 factors  including: 
renormalization scheme and scale, 
quark mass effects, higher twist, isospin violation, and PDF uncertainties. 

To simplify this analysis, we first 
examine the influence of these factors on the
LO expression:
$\Delta xF_3 = 4 x\{ s(\xi,\mu) - c(x,\mu) \}$; 
after using this as a ``toy model,"
we will then return to the full NLO  calculation in the next 
section.\footnote{Note that a possible $s\neq {\bar s}$ asymmetry
%Kretzer2: ref added
\cite{brodsky}
would average out in  $\Delta xF_3$.}
 For most variables, the simplified LO is sufficient to display the general behavior
of the full NLO result. There are two exceptions: 
1) the scheme dependence, and 2) the PDF dependence. 
These factors depend on the interplay of both 
the quark-initiated LO contributions
as well as the NLO gluon-initiated contributions.  For this reason, we will
postpone discussion of these effects until the following section.

%%%%%%%%%%%%%%%%%%%%%%%%%%%%%%%%%%%%%%%%%%%%%%%%%%%%%%%%%%%%%%%%%%%%%%%%%%%%%%
\subsection{Charm Mass}
%%%%%%%%%%%%%%%%%%%%%%%%%%%%%%%%%%%%%%%%%%%%%%%%%%%%%%%%%%%%%%%%%%%%%%%%%%%%%%
\figmc

We start by examining the effect of the charm mass, $m_c$, on 
$\Delta xF_3$.  In \fig{mc}, we plot the leading-order expression
$\Delta xF_3 = 4 x\{ s(\xi,\mu) - c(x,\mu) \}$ 
 {\it vs.} $Q^2$ for values
of $m_c$ in the range $=[0,1.8]$ GeV using steps of 0.2 GeV.
Here, we define $\xi= x(1+m_c^2/Q^2)$ which is a ``slow-rescaling"
type of correction\cite{slowres,acot} which (crudely) includes mass
effects by shifting the $x$ variable.
``Slow rescaling'' naturally arises at LO from the single-charm
threshold condition($W>m_c$) and is required at 
NLO for consistent mass factorization \cite{acot,gkr1}.

Note, the result of this
correction is most significant at  low $Q^2$.
To isolate the kinematic $m_c$ dependence, we have chosen the scale 
$\mu^2=Q^2+1.6^2 \, GeV^2$. 
(We will separately vary the scale in the following subsection.)

Note that this exercise is only altering the charm mass in one aspect of the 
calculation; to be entirely consistent it would be necessary to obtain 
parton distributions (particularly the charm quark) using fits with different 
charm masses. 
For charm masses in the range 1.2 to 1.8 GeV, this will be a small effect;
for charm masses below $\sim 1.2$ GeV, 
(which is below the experimentally allowed range\cite{hera,pdg}), 
such issues become important and the curves of \fig{mc} will be modified,
i.e.\ lower $m_c$ will lead to longer evolution for charm, a larger charm
distribution, and a lowering of the curves in \fig{mc}.

While taking $m_c\to 0$ does raise the theoretical curves in the 
regions where we observe the discrepancy (namely, the low
$Q$ region), varying $m_c$ even within the wide range of $[1.2,1.8]$ GeV
(lower 4 curves) does not give us sufficient flexibility to match
either the shape or normalization of the data.

%%%%%%%%%%%%%%%%%%%%%%%%%%%%%%%%%%%%%%%%%%%%%%%%%%%%%%%%%%%%%%%%%%%%%%%%%%%%%%
\subsection{$\mu$ Scale}  \label{sec:scale}
%%%%%%%%%%%%%%%%%%%%%%%%%%%%%%%%%%%%%%%%%%%%%%%%%%%%%%%%%%%%%%%%%%%%%%%%%%%%%%
\figmu

Next, we investigate the variation of 
$\Delta xF_3 = 4 x\{ s(\xi,\mu) - c(x,\mu) \}$ with
the renormalization/factorization scale $\mu$. 
 We use three choices of the $\mu^2$ scale: 
\begin{itemize}
\item $\mu^2 = Q^2$\ , 
\item $\mu^2 = Q^2+m_c^2$\ , 
\item $\mu^2 = \ptmax^2$\ .
\end{itemize}
\noindent
Of course, $\Delta xF_3$ with $\mu^2$ scales of 
$Q^2$ and $Q^2+m_c^2$ differ only at 
lower values of $Q^2$; 
$\Delta xF_3$ with $\mu^2$ scale of 
 $\ptmax^2$ is comparable to $Q^2$ and $Q^2+m_c^2$ at
larger $x$, but lies below for smaller $x$.  
The scale choice  $Q^2+m_c^2$ leads to an 
improvment over $Q^2$ by providing a lower bound on $\mu$
to keep the scale in the perturbative region.

The choice of $\ptmax$ is motivated, in part, by some  
observations by Collins.\cite{collinsPT} 
 To display the relationship between $\ptmax$ and $Q$, in 
\fig{bazarko}  we plot both $\ptmax$ and $Q$ {\it vs.} $x$ for 4 choices of
$y$; 
note\cite{bazarko} that the $x$-dependence of  
$\ptmax$ is opposite that of $Q$.
 While tying the scale choice of $\mu$ to $\ptmax$ has some interesting 
intuitive interpretations,\footnote{Technically, 
the interpretation is in terms of the characteristic
$P_T$ of the partonic subprocess, but this is unobservable; therefore
$\ptmax$ is used instead.\cite{collinsPT}}
for small $x$ this scale clearly becomes too large
for the relevant physics. 
 In any case, it cannot help us with our $\Delta xF3$ problem as 
the scale choice of  $\ptmax$  moves the theory curves {\it away} from the data.

%%%%%%%%%%%%%%%%%%%%%%%%%%%
\figbazarko

%%%%%%%%%%%%%%%%%%%%%%%%%%%%%%%%%%%%%%%%%%%%%%%%%%%%%%%%%%%%%%%%%%%%%%%%%%%%%%
\subsection{Higher Twist}
%%%%%%%%%%%%%%%%%%%%%%%%%%%%%%%%%%%%%%%%%%%%%%%%%%%%%%%%%%%%%%%%%%%%%%%%%%%%%%
\fight

We now illustrate the potential effects due to higher twist contributions. 
We parameterize such contributions by multiplying the leading-twist
terms by a correction factor that will increase at low $Q^2$. 
 More specifically we use:\cite{thorneHT}
\begin{equation}
F_i^{HT}(x,Q^2) = F_i^{LT}(x,Q^2) 
\left(
1 + \frac{D_i(x)}{Q^2}
\right)
\end{equation}
where $i=\{1,2,3\}$.
We vary $D_3(x)$ over the range [0,10] $GeV^2$ 
in steps of $1~GeV^2$. 
(For the purposes of our simple illustration, 
it is sufficient to take $D_3(x)$ to be independent of $x$.)
We will find that this range is well beyond what is allowed by experiment; 
however,
we display this exaggerated range to make the effect of the higher twist
contributions evident. 

To normalize this choice with the allowable range consistent with data,
we compare with the MRST higher twist fit which extracted a limit on the
function $D_2(x)$.  We see from the table of $D_2(x)$ that the allowed
contribution from the higher twist terms is quite small.
In addition, we note that the sign for  $D_2(x)$ obtained  for the 
relevant small-$x$ region tends to be negative--exactly the opposite 
sign that  is needed to move the theory toward the data. 

As we have shown constraints on $D_2(x)$, the obvious question is
should we expect  $D_3(x)$ to be substantially different?
 A calculation of the power corrections using renormalons\cite{DasguptaWebber}
suggests that while both  $D_1(x)$ and  $D_3(x)$ are of the same order of 
magnitude as  $D_2(x)$ and have similar $x$-dependence, that 
 $D_1(x)$ and  $D_3(x)$ are even more negative than  $D_2(x)$;
again, this trend would move the theory farther from the data. 
Therefore, we conclude that  that  $D_2(x)$ represents a conservative 
limit for the  $D_3(x)$ power corrections.

Examining \fig{ht}, we note that it would take an enormous higher
twist contribution to bring the normalization of the curves in the
range of the data points, and still the shape of the $Q^2$-dependence
is not well matched; hence, we conclude that this is not a compelling
solution.

%%%%%%%%%%%%%%%%%%%%%%%%%%%%%%%%%%%%%%%%%%%%%%%%%%%%%%%%%%%%%%%%%%%%%%%%%%%%%%
\tableht
%%%%%%%%%%%%%%%%%%%%%%%%%%%%%%%%%%%%%%%%%%%%%%%%%%%%%%%%%%%%%%%%%%%%%%%%%%%%%%

%%%%%%%%%%%%%%%%%%%%%%%%%%%%%%%%%%%%%%%%%%%%%%%%%%%%%%%%%%%%%%%%%%%%%%%%%%%%%%
\subsection{Isospin Violations}
%%%%%%%%%%%%%%%%%%%%%%%%%%%%%%%%%%%%%%%%%%%%%%%%%%%%%%%%%%%%%%%%%%%%%%%%%%%%%%
\figisospin

The naive parton model identity in \eq{lo2} is modified if
the full (non-diagonal) CKM structure, NLO QCD radiative corrections,
and QCD based charm production are taken into account.
 This expression is also modified {\it even in leading-order} if we have a
violation of exact $p\leftrightarrow n$ isospin-symmetry, (or charge
symmetry); {\it e.g.}, $u_n(x) \not\equiv d_p(x)$.  In deriving
\eq{lo2}, isospin-symmetry was necessary to guarantee that
the $u,d$-contributions cancel out in the difference, thereby leaving
only the $s(x)$ and $c(x)$ contributions.\footnote{%
 Note we have not investigated shadowing corrections.
The CCFR data on Fe is converted with an isoscalar correction,
and the corresponding uncertainties are included in the
data.\cite{ccfrsf}
}

The validity of exact {\it charge symmetry} (CS) has recently been
reexamined.\cite{csv}  Residual $u,d$-contributions to $\Delta xF_3$
from {\it charge symmetry violation} (CSV) 
would be amplified due to enhanced valence
components $\{u_v(x),d_v(x)\}$, and because the $d\rightarrow u$
transitions are not subject to slow-rescaling corrections which
strongly suppress the $s\rightarrow c$ contribution to $\Delta xF_3$, 
as illustrated in \fig{mc}.

We will examine possible contributions to $\Delta xF_3$
by considering a ``toy'' model to parameterize the CS-violations. 
 In isospin space, we can parameterize a general transformation as
a rotation: 
\begin{eqnarray}
\label{eq:toy1}
\left|q\right>_n^{\rm CSV} &=& 
\qquad  N_q \sum_{q^\prime}
R_{q q^\prime}(\theta ) \left|q^\prime \right>_p
\end{eqnarray}
where $R$ is a rotation matrix, and $N_q$ is the 
normalization factor. 
 For example, in this model the $u$-distribution in the neutron 
would be related to the proton distributions via the relation: 
\goodbreak
\begin{eqnarray}
\label{eq:toy2}
&& u_n^{\rm CSV}(x,Q^2) = \nonumber \\ \nobreak
&& 
N_u^2 
\left[ 
  \cos^2(\theta ) u_p(x,Q^2)
+ \sin^2(\theta ) d_p(x,Q^2) 
\right]
\end{eqnarray}
For $\theta=\pi/2$, we recover the symmetric limit $u_p(x,Q^2)=d_n(x,Q^2)$.
%Kretzer2
Note that we define the normalization $N_u$ such that 
this model preserves the sum rule: 
\begin{equation}
\int dx \ u_n^{\rm CSV}(x,Q^2)-{\bar u}_n^{\rm CSV}(x,Q^2)=1
\quad .
\end{equation}
Note that \eq{toy1} should not be considered a serious
theory,\footnote{{\it E.g.}, it does strictly speaking not commute
with evolution.}  but rather a simple one-parameter ($\theta$) model
which is flexible enough to illustrate a range of CSV effects for
$\Delta xF_3$.

In \fig{isospin} we vary $\theta$ over the  its maximum  range 
$[0,\pi/2]$ in steps  of $\pi/20$.
The exact charge symmetry limit ($\theta_0 = \pi /2$) of \eq{toy2} 
corresponds to  the lowest curve in  \fig{isospin}. 
Note the effect of the CSV contribution 
monotonically increases  as $\theta$ deviates from the 
charge symmetry (CS) limit $\theta_0=\pi/2$. 
From the plot we observe that 
a violation of $\theta_0 - \theta \gtrsim (3/10)(\pi/2)$ is required
to bring the theory in the neighborhood of the data. 

 At the relevant $x$ values of interest for $\Delta xF_3$,
this translates  (via  \eq{toy2}  and its analogue for $d_n$)
into a $\sim 20\%$ symmetry violation for 
$u_n$,  and a $\sim 10\%$ symmetry violation for $d_n$. 
Specifically, 
\begin{eqnarray}
\frac{d_p(x)}{u_n^{\rm CSV}(x)} &\simeq& 1.2 \\
\frac{u_p(x)}{d_n^{\rm CSV}(x)} &\simeq& 0.9  \quad .
\end{eqnarray}
Since the  electroweak couplings are flavor-independent, 
($V_q=A_q=1 \, \forall \,  q$),  $\Delta xF_3$ is in principle
insensitive to a re-shuffling of the CSV contributions 
between $u_n$ and $d_n$.
 In particular, if we define the shift  due to CSV in  $\Delta xF_3$ as  
$\delta\Delta xF_3$, we find:
\begin{equation}
\delta\Delta xF_3 \equiv \frac{x}{4} 
( \delta d + \delta \bar{d} - \delta u - \delta \bar{u} )
\label{eq:Delta}
\end{equation}  
where 
\begin{equation}
\delta u (x,Q^2) \equiv u_n^{\rm CSV}(x,Q^2)- d_{p}(x,Q^2)
\quad .
\label{eq:delta}
\end{equation}
The expression for $\delta\Delta xF_3(x,Q^2)$ of \eq{Delta} is
evaluated at $Q^2=4\, {\rm GeV}^2$ from \eq{toy2} with
$\theta_0 -\theta  = (3/10) (\pi /2)$ and is plotted in 
\fig{iso}.\footnote{
Again, the detailed $x$-shape from \eq{toy2} 
should {\it not} be taken as a serious model prediction.}

%%%%%%%%%%%%%%%%%%%%%%%%%%%%%%%%%%%%%
\figiso

Although this level of isospin violation certainly improves the description 
of $\Delta xF_3$, it is necessary to consider precisely what level of violation
is actually allowed by other experimental data. For instance, it has previously
been suggested that the discrepancy between $F_2$ from neutrino and muon 
data itself may be due to isospin violation.\cite{csv} This type of violation
required that $\bar u$ be very different to $\bar d$ in the region of 
interest. 

The measurement of the lepton charge asymmetry in W decays from the
Tevatron \cite{cdfWasym,bflmy} places tight constraints on the up and down quark
distributions in the range $0.007 < x < 0.24$, 
constraining them to be approximately as specified in the parton sets 
obtained by the global analyses. While only strictly telling 
us about parton distributions in the proton, this data rules out an 
isospin violation of this type to about $5\%$ as demonstrated in \cite{bflmy}.  

However, there are other strong constraints on isospin violation. 
For example, we note that while the toy model above leaves the neutron singlet
combination $q + {\bar q}$ invariant at the $\lesssim 2\%$ level in
the region $x\, \epsilon \, [0.01;0.1]$,
 it would lower the NC observable: 
\begin{equation}
\label{eq:ncobs}
\left. 
 \left[
  \frac{4}{9} (u+{\bar u}) 
+ \frac{1}{9}  (d+{\bar d})
 \right]_n \  
\right|_{x\epsilon [0.01;0.1]}
\end{equation}
by about 10\%.
An effect of this size would definitely
be visible in NMC $F_2^n/F_2^p$ data
which has an uncertainty of order a few percent 
in this kinematic region, and acts as a major constraint.\cite{nmc}

At this point, one could play clever games to evade the constraints of 
specific experiments. 
For example, a  re-shuffling of CSV contributions  between 
the individual $\delta q, \delta {\bar q}$ in \eq{Delta}
according to:
\begin{eqnarray}
&x u_n&=x d_p - \frac{2}{5}  \delta\Delta xF_3 
\\ \nonumber
&x d_n&=x u_p + \frac{8}{5}   \delta\Delta xF_3 
\\ \nonumber
&x{\bar q}_n&: {\mbox{\rm analogous}}
\end{eqnarray}
would keep \eq{ncobs} invariant. However, this  would in turn 
raise the CC observable \eq{Delta}
by $\gtrsim 5\%$; though it would help to explain the excess in 
$\delta xF_3$, it would spoil the new-found compatibility between neutral
current and charged current data.

In addition, there are also fixed-target Drell-Yan 
experiments\cite{na51,e866} such as
NA51 and E866 which precisely measure $\bar{d}/\bar{u}$
in the range\cite{e866} $0.04 < x < 0.27$, and are also sensitive to isospin
violating effects. 
 
We therefore conclude that the many precise data sets which constrain
different combinations of the PDF's probably leave no room for CSV 
contributions of the magnitude necessary to fully align the theory curves with 
the $\Delta xF_3$ data. 
However, we should add the caveat that 
an exhaustive investigation of the interplay of these different data sets
and their influence on   $\Delta xF_3$ 
will only be possible within a global PDF analysis

%%%%%%%%%%%%%%%%%%%%%%%%%%%%%%%%%%%%%%%%%%%%%%%%%%%%%%%%%%%%%%%%%%%%%%%%%%%%%%
\section{NLO calculation of $\Delta xF_3/2$}
%%%%%%%%%%%%%%%%%%%%%%%%%%%%%%%%%%%%%%%%%%%%%%%%%%%%%%%%%%%%%%%%%%%%%%%%%%%%%%
\label{sec:NLO}

Now that we have used the leading-order expression for  $\Delta xF_3$
to systematically investigate dependence of this observable on 
various parameters, in this section we 
now turn to the  full NLO calculation.

%%%%%%%%%%%%%%%%%%%%%%%%%%%%%%%%%%%%%%%%%%%%%%%%%%%%%%%%%%%%%%%%%%%%%%%%%%%%%%
\subsection{Contributions to the NLO Calculation} 
%%%%%%%%%%%%%%%%%%%%%%%%%%%%%%%%%%%%%%%%%%%%%%%%%%%%%%%%%%%%%%%%%%%%%%%%%%%%%%
\label{sec:contributions}
\figxfnlo

 In \fig{xfnlo}, we have plotted the 
LO and NLO calculations for  $\Delta x F_3$ {\it vs.} $Q^2$ 
on an isoscalar target
in 3 and 4 flavor schemes.
 The 3-flavor LO calculation (f=3, $\alpha_s^0$) involves 
primarily the strange quark contribution, $s(x)$, as the charm distribution
is excluded in this case. 
 When the higher order 
terms are included  (f=3, $\alpha_s^1$), this result moves (substantially) 
 toward  the predictions of the 4-flavor scheme.
 We note that while  the 3-flavor LO calculation (f=3, $\alpha_s^0$)
  appears consistent with the data, 
we cannot take this result as a precise theoretical prediction as this
simplistic result is highly dependent on scheme and scale choices; a result
that is verified by the large shift in going from LO to NLO. 

The pair of curves in the 4-flavor scheme  
(using the CTEQ4HQ distributions) nicely illustrates
how the charm distribution $c(x,\mu^2)$ evolves as $\ln(Q^2/m_c^2)$ for
increasing $Q^2$; note, $c(x,\mu^2)$ enters with a negative sign 
so that the 4-flavor result is below the 3-flavor curve. 
For the scale choice, we take $\mu=\sqrt{Q^2+m_c^2}$.
While the scale choice $\mu=Q$ is useful for instructive purposes
such as demonstrating the matching of the 3- and 4-flavor calculations
at $\mu=Q=m_c$, the choice $\mu=\sqrt{Q^2+m_c^2}$ is more practical
as it provides a lower bound on $\mu$ which is important for 
the PDF's and $\alpha_s(\mu)$. 
({\it Cf.}, \sec{schemes}, and Ref.~\cite{schmidt}.)

Additionally we note the stability of the 4-flavor scheme 
in contrast to the 3-flavor scheme. The shift of the curves 
when including the NLO contributions is quite minimal, particularly
when compared with the  3-flavor result.\cite{buza,csn}
 This suggests that organizing the calculation to include the charm quark
as a proton constituent can be advantageous even at relatively low values
of the energy scale.

%%%%%%%%%%%%%%%%%%%%%%%%%%%%%%%%%%%%%%%%%%%%%%%%%%%%%%%%%%%%%%%%%%%%%%%%%%%%%%
\subsection{PDF Uncertainties: s(x), ...}  \label{sec:squark}
%%%%%%%%%%%%%%%%%%%%%%%%%%%%%%%%%%%%%%%%%%%%%%%%%%%%%%%%%%%%%%%%%%%%%%%%%%%%%%
\figpdf

In \fig{pdf}, we show the variation of NLO calculation of $\Delta
xF_3$ on the strange-quark PDF.  To obtain a realistic assessment of
the $s(x)$ dependence, we have use the NLO calculation with PDF's
based on the MRST set which are re-fit with the value of $\kappa=2
s/(\bar{u}+\bar{d})$ constrained to be $\kappa=\{0.50, 0.78, 1.00 \}$.
Note, by re-fitting the PDF's with the chosen value of $\kappa$ we are
assured to have an internally consistent set of PDF's with appropriate
matching between the quarks and gluon, and with the sum-rules
satisfied.\footnote{%
Note, $\kappa$ is certainly $Q$-dependent, and the values for $\kappa$
quoted above correspond to the $Q_0$ of the evolution.  
While $\kappa$ compares the integral of $s(x)$ to the seq-quarks, 
there is also the possibility of an $x$-dependent variation. 
This has been studied in the fits of the 
strange-sea\cite{ccfrlo,bazarko,nutev,bpz}; we shall
find that such subtle effects can play no role in resolving the 
$\Delta xF_3$ issue. 
}

%%%%%%%%%%%%%%%%%%%%%%%%%%%%%%%%%%%%%%%%%%%%%%%%%%%%%%%%%%%%%%%%%%%%%%%%%%%%%%
\tablek
%%%%%%%%%%%%%%%%%%%%%%%%%%%%%%%%%%%%%%%%%%%%%%%%%%%%%%%%%%%%%%%%%%%%%%%%%%%%%%

The choice $\kappa=0.50$ is in line with the many experimental
determinations of $\kappa$, {\it cf.}, \tab{kappa}; as expected, this
prediction lies farthest from the data points.

The choice $\kappa=0.78$ is taken as an extreme upper limit given the
experimental constraints; actually, in light of the results of
\tab{kappa}, this is arguably {\it beyond} present experimental
bounds. This prediction is marginally consistent at the outer reach of
the systematic + statistical error bars.

Finally, we take an $SU(3)$ symmetric set ($\kappa=1$) purely for
illustrative purposes. It is interesting to note that even this
extreme value is still below the central value of the data points at
the higher $x$ values.

In conclusion we note that increasing the strange quark distribution
does succeed in moving the theory toward the data; however, our 
consistent NLO analysis presented here suggests that the we have 
only limited freedom to increase $s(x)$, and that this alone is not 
sufficient to obtain good agreement between theory and data.

%%%%%%%%%%%%%%%%%%%%%%%%%%%%%%%%%%%%%%%%%%%%%%%%%%%%%%%%%%%%%%%%%%%%%%%%%%%%%%
\subsection{Scheme Choice} 
%%%%%%%%%%%%%%%%%%%%%%%%%%%%%%%%%%%%%%%%%%%%%%%%%%%%%%%%%%%%%%%%%%%%%%%%%%%%%%
%%% Kretzer: label introduced
\label{sec:schemes}
\figscheme

In our final section, we present the best theoretical predictions
presently available to demonstrate the scheme dependence of $\Delta xF_3(x,Q)$.
Specifically, in \fig{scheme} we show  predictions for:
\begin{itemize}

\item  NLO FFS GRV.\cite{grv98}

\item  NLO VFS TR calculation.\cite{tr}

\item  NLO VFS ACOT calculation with CTEQ4 PDF's.\cite{acot,kretzer,cteq4}

\item  NLO VFS ACOT calculation with CTEQ5 PDF's.\cite{acot,kretzer,cteq5}

\end{itemize}
All these calculations use NLO matrix elements, and are matched
with appropriate global PDF's which are fitted in the proper scheme. 

The first observation we make is how closely these four predictions match, 
especially given the wide variation displayed in previous 
plots such as \fig{xfnlo}. 
 In hindsight, this result is simply a consequence of the fact that  while
different renormalization schemes can produce different results, 
this difference can only be higher order.\footnote{%
To be precise, different renormalization schemes can differ by
i) terms of higher order in the perturbation series, and
ii) terms of higher twist which do not factorize.\cite{collins98,sacot}}
Thus, the difference between these curves is indicative of terms of
order $\alpha_s^2$ which have yet to be calculated.\footnote{%
For asymptotic results at order  $\alpha_s^2$, see Ref.~\cite{buza}.}
When  terms of order  $\alpha_s^N$ are
included, the span of these predictions will be systematically 
reduced to order  $\alpha_s^{N+1}$.

In  \fig{scheme}, we note the very close agreement among the VFS 
calculations, particularly the TR calculation 
and the ACOT  calculation with CTEQ4 PDF's.
 The  ACOT  calculation with the two CTEQ curves show primarily the 
effect of the charm distribution, as CTEQ4 uses $m_c=1.6$ and 
 CTEQ5 uses $m_c=1.3$.
The  GRV calculation shows the effect of using yet a different scheme,
in this case  a FFN scheme, with its appropriately matched PDF.
Were we to use MRST or CTEQ PDF's, the spread of these theory curves
would decrease; however, this would most likely represent an 
underestimate of the true theoretical uncertainties arising from 
both the hard cross section and PDF's.\footnote{%
The computation of PDF errors is a complex subject. 
For some recent approaches to this
topic see:  Refs.\cite{MRST,cteq5,pdferrors,bpz}.}

While we consider it a triumph of QCD that different schemes truly 
yield comparable results (higher order terms aside), we should be cautious 
and note that the spread of these curves can only underestimate 
the true theoretical uncertainty.  Note that GRV has a rather different
strange distribution due to a different philosophy of obtaining this
distribution rather than due to a different scheme.

%%%%%%%%%%%%%%%%%%%%%%%%%%%%%%%%%%%%%%%%%%%%%%%%%%%%%%%%%%%%%%%%%%%%%%%%%%%%%%
\section{Conclusions and Outlook}
%%%%%%%%%%%%%%%%%%%%%%%%%%%%%%%%%%%%%%%%%%%%%%%%%%%%%%%%%%%%%%%%%%%%%%%%%%%%%%

 Comprehensive analysis of the neutrino data sets can provide incisive
 tests of the theoretical methods, particularly in the low $Q^2$
 regime, and enable precise predictions that will facilitate new
 particle searches by constraining the PDF's.  This document serves as
 a progress report, and work on these topics will continue in the
 future.

Theoretical predictions for $\Delta xF_3$ undershoot preliminary fixed
target data at the $\sim 1\, \sigma$-level at low $x$ and $Q$.  The
neutrino structure function $\Delta xF_3$ is obviously 
sensitive\cite{bag} to the strange sea of the nucleon and the details of deep
inelastic charm production.  A closer inspection reveals, however,
considerable dependence upon factors such as the charm mass,
factorization scale, higher twists, contributions from longitudinal
$W^\pm$ polarization states, nuclear shadowing, charge symmetry
violation, and the  PDF's.  This makes
$\Delta xF_3(x,Q^2)$ an excellent tool to probe both perturbative and
non-perturbative QCD.

We have explored the variation of $\Delta xF_3(x,Q^2)$ on the above
 factors and found none of these to be capable of resolving the
 discrepancy between the data and theory.

Although we have not eliminated the possibility that the entire set of
parameters conspires to align the theory with the data, we have
demonstrated this to be extremely unlikely.  Of course, a definitive
answer can only be obtained by a global analysis which combines the 
neutrino data for dimuons, 
$\Delta xF_3$, $F_2^{\nu}(PMI)$, and $F_2^{\mu,e}$.

As the situation stands now, this $\Delta xF_3(x,Q^2)$ puzzle poses an
important challenge to our understanding of QCD and the related
nuclear processes in an important kinematic region. The resolution of
this puzzle is important for future data analysis, and the solution is
sure to be enlightening, and allow us to expand the applicable regime
of the QCD theory.

%%%%%%%%%%%%%%%%%%%%%%%%%%%%%%%%%%%%%%%%%%%%%%%%%%%%%%%%%%%%%%%%%%%%%%%%%%%%%%
\section*{Acknowledgments} 
%%%%%%%%%%%%%%%%%%%%%%%%%%%%%%%%%%%%%%%%%%%%%%%%%%%%%%%%%%%%%%%%%%%%%%%%%%%%%%

 This work is supported
by the Royal Society,
the U.S. Department of Energy, 
the National Science Foundation, 
the Lightner-Sams Foundation,
and
 the `Bundesministerium
f\"{u}r Bildung, Wissenschaft, Forschung und Technologie', Bonn.

We thank
J.~Bluemlein,
A.~Bodek,
J.~Conrad, 
J.~Morfin,
S.~Kuhlmann,
R.G.~Roberts,
H.~Schellman,
M.~Shaevitz,
J.~Smith, 
and
W.-K.~Tung,
 for valuable discussions.

%%%%%%%%%%%%%%%%%%%%%%%%%%%%%%%%%%%%%%%%%%%%%%%%%%%%%%%%%%%%%%%%%%%%%%%%%%%%%%
%\begin{thebibliography}{99}

%%%%%%%%%%%%%%%%%%%%%%%%%%%%%%%%%%%%%%%%%%%%%%%%%%%%%%%%%%%%%%%%%%%%%%%%%%%%%%

%%%%%%%%%%%%%%%%%%%%%%%%%%%%%%%%%%%%%%%%%%%%%%%%%%%%%%%%%%%%%%%%%%%%%%%%%%%%%%
\end{document}